\documentclass[preprint]{ptephy}

\preprintnumber{KYUSHU-HET-212, KEK-TH-2216}

\usepackage{amssymb}
\usepackage{amsthm}
\usepackage{amsmath}
\usepackage{booktabs}
\usepackage{bbm}
\usepackage{bm}
\usepackage{mathtools}
\usepackage{subcaption}

\DeclareMathOperator{\Real}{Re}
\DeclareMathOperator{\tr}{tr}

\let\Im\relax
\DeclareMathOperator{\Im}{Im}

\DeclareMathOperator{\Li}{Li}

\numberwithin{equation}{section}

\begin{document}

\title{More on the infrared renormalon in $SU(N)$ QCD(adj.) on $\mathbb{R}^3\times S^1$}

\author{%
\name{\fname{Masahiro} \surname{Ashie}}{1},
\name{\fname{Okuto} \surname{Morikawa}}{1},
\name{\fname{Hiroshi} \surname{Suzuki}}{1,\ast}, and
\name{\fname{Hiromasa} \surname{Takaura}}{2}
}

\address{%
\affil{1}{Department of Physics, Kyushu University, 744 Motooka, Nishi-ku,
Fukuoka 819-0395, Japan}
\affil{2}{Theory Center, High Energy Accelerator Research Organization (KEK),
Tsukuba, Ibaraki 305-0801, Japan}
\email{hsuzuki@phys.kyushu-u.ac.jp}
}

\date{\today}

\begin{abstract}
We present additional observations to previous studies on the infrared (IR)
renormalon in $SU(N)$ QCD(adj.), the $SU(N)$ gauge theory with $n_W$-flavor
adjoint Weyl fermions on~$\mathbb{R}^3\times S^1$ with the $\mathbb{Z}_N$
twisted boundary condition. First, we show that, for arbitrary finite~$N$, a
logarithmic factor in the vacuum polarization of the ``photon'' (the gauge
boson associated with the Cartan generators of~$SU(N)$) disappears under the
$S^1$~compactification. Since the IR renormalon is attributed to the presence
of this logarithmic factor, it is concluded that there is no IR renormalon in
this system with finite~$N$. This result generalizes the observation made by
Anber and~Sulejmanpasic [J. High Energy Phys.\ \textbf{1501}, 139 (2015)] for
$N=2$ and~$3$ to arbitrary finite~$N$. Next, we point out that, although
renormalon ambiguities do not appear through the Borel procedure in this
system, an ambiguity appears in an alternative resummation procedure in which a
resummed quantity is given by a momentum integration where the inverse of the
vacuum polarization is included as the integrand. Such an ambiguity is caused
by a simple zero at non-zero momentum of the vacuum polarization. Under the
decompactification~$R\to\infty$, where $R$ is the radius of the $S^1$, this
ambiguity in the momentum integration smoothly reduces to the IR renormalon
ambiguity in~$\mathbb{R}^4$. We term this ambiguity in the momentum integration
``renormalon precursor''. The emergence of the IR renormalon ambiguity
in~$\mathbb{R}^4$ under the decompactification can be naturally understood with
this notion.
\end{abstract}

\subjectindex{B00, B06, B32}
\maketitle

\section{Introduction}
\label{sec:1}
Perturbative expansion of observables typically gives divergent asymptotic
series. Such divergent behavior is caused by factorial growth of perturbative
coefficients and often induces intrinsic errors in perturbative predictions.
One of the sources of factorial growth is known as the
renormalon~\cite{tHooft:1977xjm,Beneke:1998ui}. This is closely related to
renormalization properties, and in asymptotically free theories the infrared
(IR) renormalon gives inevitable uncertainties in perturbation theory. The fate
of the IR renormalon, for instance how its ambiguity is eliminated, has not
been well understood so far.

In~Refs.~\cite{Argyres:2012vv,Argyres:2012ka,Dunne:2012ae,Dunne:2012zk}, a
conjecture concerning the IR renormalon was proposed: In an $S^1$~compactified
spacetime with the $\mathbb{Z}_N$ twisted boundary condition, the ambiguity
associated with the IR renormalon is canceled against the ambiguity associated
with the integration of quasi-collective coordinates of a semi-classical
quasi-solution called a bion~\cite{Unsal:2007jx}. This conjecture suggests a
semi-classical picture of the IR renormalon in an analogous manner to the
cancellation of ambiguities between the proliferation of Feynman diagrams and
the instanton--anti-instanton pair~\cite{Bogomolny:1980ur,ZinnJustin:1981dx}.
The suggested structure would be fascinating to the resurgence program in
asymptotically free field theories~\cite{Dunne:2015eaa}. To examine this
conjecture, a study of IR renormalons in theories on the $S^1$~compactified
spacetime was performed~\cite{Anber:2014sda,Ishikawa:2019tnw,Ashie:2019cmy,%
Ishikawa:2019oga,Ishikawa:2020eht}.
See~Refs.~\cite{Fujimori:2016ljw,Fujimori:2018kqp} for detailed analyses from
the bion side.

Contrary to the conjecture, however, it was argued in~Ref.~\cite{Anber:2014sda}
that the bion ambiguity does not correspond to renormalon ambiguities, because
it was found that the IR renormalon is absent in the $SU(N)$ QCD (adj.)
for~$N=2$ and~$3$, in which the bion ambiguity exists; this system is defined
as the $SU(N)$ gauge theory with $n_W$-flavor adjoint Weyl fermions with the
$Z_N$ twisted boundary condition~\cite{Kovtun:2007py,Unsal:2007vu,%
Unsal:2007jx,Shifman:2008ja,Unsal:2008ch,Shifman:2009tp,Anber:2011de,%
Unsal:2012zj,Poppitz:2012sw,Poppitz:2012nz,Basar:2013sza,Poppitz:2013zqa,%
Anber:2013doa,Cherman:2014ofa,Misumi:2014raa,Anber:2014lba,Dunne:2016nmc,%
Cherman:2016hcd,Sulejmanpasic:2016llc,Yamazaki:2017ulc,Aitken:2017ayq,%
Tanizaki:2017qhf} on~$\mathbb{R}^3\times S^1$. Quite recently, the perturbative
ambiguity to be canceled against the bion ambiguity has been identified
in~Ref.~\cite{Morikawa:2020agf}; it has been clarified that such a perturbative
ambiguity is not caused by the IR renormalon but by the proliferation of
Feynman diagrams and the enhancement of the amplitude of each diagram, which is
specific to the $S^1$~compactification and the twisted boundary condition. In
this way, the recent controversial issue of whether bion ambiguities truly
correspond to renormalon ambiguities or not has been settled to our
understanding.

In this paper, nevertheless, we further investigate renormalon ambiguities of a
theory on the $S^1$~compactified spacetime. This aims at understanding issues
remaining unclear about the renormalon structure in a compactified spacetime
itself. A particular purpose of this paper is to understand the relation
between two results given in~Refs.~\cite{Anber:2014sda}
and~\cite{Ashie:2019cmy} on the IR renormalon in the $SU(N)$ QCD (adj.)
on~$\mathbb{R}^3\times S^1$. In~Ref.~\cite{Anber:2014sda}, the vacuum
polarization of the ``photon'' (the gauge boson associated with Cartan
generators of~$SU(N)$) was analyzed in great detail and it was found that a
logarithmic factor in the vacuum polarization, which is responsible for the
existence of the IR renormalon, disappears as an effect of the
$S^1$~compactification. This analysis was performed explicitly for $N=2$
and~$3$ and indicates the absence of the IR renormalon. On the other hand,
in~Ref.~\cite{Ashie:2019cmy}, it was concluded that there exists an IR
renormalon for~$N=\infty$. Therefore, it is of great interest to know how the
existence of the IR renormalon depends on the value of~$N$. In the analyses of
the present paper, we rely entirely on the large-$\beta_0$
approximation~\cite{tHooft:1977xjm,Beneke:1994qe,Broadhurst:1993ru,%
Ball:1995ni}, which is a somewhat ad hoc but widely used approximation in the
study of renormalons in asymptotically free theories (see below); this
approximation was also adopted in~Ref.~\cite{Ashie:2019cmy}.

In the first part of this paper, we show that the logarithmic factor in the
vacuum polarization of the photon disappears for arbitrary finite~$N$, by
employing expressions obtained in~Ref.~\cite{Ashie:2019cmy}. This result
generalizes the result in~Ref.~\cite{Anber:2014sda}, which studied the cases
with $N=2$ and $3$.\footnote{Our treatment of the gauge field loop diagrams is
somewhat different from that in~Ref.~\cite{Anber:2014sda}; see below.} We
conclude that the IR renormalon does not exist for \emph{any\/} finite~$N$. We
also make remarks on how the point $N=\infty$ should be regarded as singular.

The absence of the IR renormalon for arbitrary finite~$N$ is, however, somewhat
peculiar because it indicates that the IR renormalon does not exist regardless
of the details of the theory as long as the $S^1$~compactification is
considered. On the other hand, we know that the IR renormalon indeed exists
in~$\mathbb{R}^4$. Then, the question arises of how the IR renormalon
in~$\mathbb{R}^4$ can emerge in the decompactification limit starting from the
theory on the $S^1$~compactified spacetime.\footnote{In most of this theory,
$R$ dependence is controlled by the combination~$NR$ instead of~$R$ due to the
twisted boundary condition. Then, it is naively expected that as~$N$ becomes
larger the theory becomes equivalent to that
on~$\mathbb{R}^4$~\cite{Gross:1982at}.} To gain an insight to this issue, in
the second part of this paper, we point out that although renormalon
ambiguities do not appear through the Borel procedure, an ambiguity appears in
an alternative resummation procedure in which a resummed quantity is given by a
momentum integration where the inverse of the vacuum polarization is included
as the integrand. Such an ambiguity is caused by a simple zero at non-zero
momentum of the vacuum polarization. This ambiguity is generally different from
ordinary renormalon ambiguities, which we encounter in the Borel procedure. One
advantage of considering such an ambiguity is that we can naturally understand
how the IR renormalon ambiguity in~$\mathbb{R}^4$ emerges under the
decompactification limit~$R\to\infty$. We term this ambiguity in the momentum
integration the ``renormalon precursor''. This ambiguity is not the IR
renormalon in the sense that it is not associated with the factorial growth of
the perturbative coefficients; it is nevertheless a ``precursor'' of the IR
renormalon in the sense that under the decompactification $R\to\infty$, the
renormalon precursor smoothly reduces to the IR renormalon in~$\mathbb{R}^4$.

This paper is organized as follows. In~Sect.~\ref{sec:2}, we give a review of
the IR renormalon in~$SU(N)$ QCD(adj.) and collect the necessary results
obtained in~Ref.~\cite{Ashie:2019cmy}. In~Sect.~\ref{sec:3}, we show the
absence of the logarithm factor in the vacuum polarization at the low-momentum
limit for arbitrary finite~$N$. In Sect.~\ref{sec:4}, we introduce the
``renormalon precursor'' and discuss perturbative ambiguities in the
decompactification limit. Section~\ref{sec:5} is devoted to the conclusion.
In~Appendix~\ref{sec:A}, we give a rigorous proof of the asymptotic behavior of
the vacuum polarization at the low-momentum limit. In~Appendix~\ref{sec:B}, we
present some examples to which the notion of the renormalon precursor applies;
we see that even the shift of the Borel singularity by~$-1/2$ under the
compactification~$\mathbb{R}^d\to\mathbb{R}^{d-1}\times S^1$ in some
models~\cite{Ishikawa:2019tnw,Ishikawa:2019oga} can be naturally understood by
this notion.

\section{Preparation: Basics on IR renormalon in QCD(adj.)}
\label{sec:2}
Let us start with recalling how the IR renormalon arises in QCD(adj.) in the
uncompactified spacetime~$\mathbb{R}^4$. Throughout this paper, we rely on the
large-$\beta_0$ approximation~\cite{tHooft:1977xjm,Beneke:1994qe,%
Broadhurst:1993ru,Ball:1995ni}, which extracts a certain (gauge-invariant)
sub-contribution of Feynman diagrams. For this, one first considers the
large-flavor limit~$n_W\to\infty$ with the combination~$g^2n_W$ kept fixed,
where $g$ is the gauge coupling constant. In this limit, the gauge field
propagator is dominated by the chain of the fermion one-loop vacuum bubbles.
Then, to partially incorporate the effect of the gauge field loops, the number
of flavors~$n_W$ is replaced by hand with the one-loop coefficient of the beta
function of the 't~Hooft coupling~$\lambda=g^2N$ as
\begin{equation}
   -\frac{2}{3}n_W\to\beta_0\equiv\frac{11}{3}-\frac{2}{3}n_W.
\label{eq:(2.1)}
\end{equation}
In this large-$\beta_0$ approximation, the gauge field propagator is given by
\begin{align}
   &\left\langle A_\mu^a(x)A_\nu^b(y)\right\rangle
\notag\\
   &=\frac{\lambda}{N}\delta^{ab}\int\frac{d^4p}{(2\pi)^4}
   e^{ip(x-y)}
   \frac{1}{(p^2)^2}
   \left\{
   \left[1
   -\frac{\beta_0\lambda}{16\pi^2}\ln\left(\frac{e^{5/3}\mu^2}{p^2}\right)
   \right]^{-1}
   (p^2\delta_{\mu\nu}-p_\mu p_\nu)
   +\frac{1}{\xi}p_\mu p_\nu
   \right\},
\label{eq:(2.2)}
\end{align}
where $\lambda$ is the renormalized 't~Hooft coupling in the
$\overline{\text{MS}}$ scheme at the renormalization scale $\mu$; $\xi$ is the
renormalized gauge parameter. We note that this form is actually consistent
with a renormalization group equation. From the geometric series expansion of
this expression, the perturbative expansion of a gauge-invariant physical
quantity~$\mathcal{F}(\lambda)$ is expected to have the form\footnote{We will
explicitly see such an example in~Eqs.~\eqref{eq:(4.1)}, \eqref{eq:(4.8)},
and~\eqref{eq:(4.9)}.}
\begin{equation}
   \mathcal{F}(\lambda)
   \sim\lambda\sum_{k=0}^\infty f_k
   \left(\frac{\beta_0\lambda}{16\pi^2}\right)^k,\qquad
   f_k=\int\frac{d^4p}{(2\pi)^4}\,(p^2)^\alpha
   \left[\ln\left(\frac{e^{5/3}\mu^2}{p^2}\right)\right]^k.
\label{eq:(2.3)}
\end{equation}
Here, we assume that $\alpha+2>0$ so that the perturbative expansion
of~$\mathcal{F}(\lambda)$ does not suffer from IR divergences. For $k\gg1$, the
momentum integral for~$f_k$ is dominated by the contribution of the saddle
point~$p^2=e^{5/3}\mu^2e^{-k/(\alpha+2)}$ and the large-order behavior is given by
\begin{equation}
   f_k\stackrel{k\gg1}{\sim}\frac{(e^{5/3}\mu^2)^{\alpha+2}}{16\pi^2}
   \frac{k!}{(\alpha+2)^{k+1}}.
\label{eq:(2.4)}
\end{equation}
For the Borel transform defined by
\begin{equation}
   B[\mathcal{F}](u)
   \equiv\sum_{k=0}^\infty\frac{f_k}{k!}u^k,
\label{eq:(2.5)}
\end{equation}
the above factorial growth of the perturbative coefficient~$f_k$ produces a
pole singularity at~$u=\alpha+2$:
\begin{equation}
   \frac{(e^{5/3}\mu^2)^{\alpha+2}}{16\pi^2}\frac{1}{\alpha+2-u}.
\label{eq:(2.6)}
\end{equation}
In the Borel procedure, which allows us to resum divergent series, the Borel
integral
\begin{equation}
   \frac{16\pi^2}{\beta_0}\int_0^\infty du\,B[\mathcal{F}](u)\,
   e^{-16\pi^2u/(\beta_0\lambda)}
\label{eq:(2.7)}
\end{equation}
formally gives the original quantity $\mathcal{F}(\lambda)$. However, the Borel
integral along the positive $u$-axis should be regularized due to the pole
singularity of the Borel transform at~$u=\alpha+2>0$. The integration contour
is often deformed in the complex $u$-plane
as~$\int_0^\infty\to\int_{0\pm i\delta}^{\infty\pm i\delta}$ with a small
parameter~$\delta$. Accordingly, it possesses an imaginary part, regarded as
the ambiguity associated with the pole singularity,
\begin{equation}
   \pm\pi i\frac{1}{\beta_0}(e^{5/3}\Lambda^2)^{\alpha+2},
\label{eq:(2.8)}
\end{equation}
where $\Lambda$ is the one-loop dynamical scale:
\begin{equation}
   \Lambda^2\equiv\mu^2e^{-16\pi^2/(\beta_0\lambda)}.
\label{eq:(2.9)}
\end{equation}
Equation~\eqref{eq:(2.8)} is an IR renormalon ambiguity. Since the factorial
growth of~$f_k$ comes from the momentum integration
around~$p^2=e^{5/3}\mu^2e^{-k/(\alpha+2)}$, which goes to~$0$ as~$k\to\infty$
(i.e., for the large-order behavior of perturbation theory), the persistent
presence of the logarithmic factor~$\ln p^2$ in the vacuum polarization
in~Eq.~\eqref{eq:(2.2)} toward~$p^2=0$ is crucial for the existence of the IR
renormalon. For example, if the vacuum polarization approaches a constant
as~$p^2\to0$, we do not have factorial growth of perturbative coefficients.

Now, we explain how the expression~\eqref{eq:(2.2)} is modified under the
$S^1$~compactification, $\mathbb{R}^4\to\mathbb{R}^3\times S^1$ based
on~Ref.~\cite{Ashie:2019cmy}. We compactify the $x_3$-direction and impose the
$\mathbb{Z}_N$ twisted boundary condition along~$S^1$ (see~Eqs.~(2.3)--(2.7)
in~Ref.~\cite{Ashie:2019cmy} for a detailed definition). Since the twisted
boundary condition is expressed in terms of Cartan generators of~$SU(N)$, it is
convenient to decompose the field in the Cartan--Weyl basis as
\begin{align}
   A_\mu(x)&=-i\sum_{\ell=1}^{N-1}A_\mu^\ell(x)H_\ell
   -i\sum_{m\neq n}A_\mu^{mn}(x)E_{mn},
\label{eq:(2.10)}
\end{align}
where $H_\ell$ are Cartan generators and $E_{mn}$ are root generators
of~$SU(N)$. In what follows, we refer to the Cartan components $A_\mu^\ell(x)$
as the ``photon'', and the root components $A_\mu^{mn}(x)$ as the ``W-boson'';
they have rather different properties.  

Gauge field propagators are given in Eq.~(2.37) of~Ref.~\cite{Ashie:2019cmy}
in the large-$\beta_0$ approximation as\footnote{In this paper, we write the
fields subject to the twisted boundary condition simply without putting the
tilde~$\tilde{\phantom{m}}$, unlike Ref.~\cite{Ashie:2019cmy}.}
\begin{align}
   &\left\langle A_\mu^\ell(x)A_\nu^r(y)\right\rangle
\notag\\
   &=\frac{\lambda}{N}\int\frac{d^3p}{(2\pi)^3}
   \frac{1}{2\pi R}\sum_{p_3}
\notag\\
   &\qquad\qquad{}
   \times e^{ip(x-y)}
   \frac{1}{(p^2)^2}
   \left\{
   \left[(1-L)^{-1}\right]_{\ell r}p^2\mathcal{P}_{\mu\nu}^L
   +\left[(1-T)^{-1}\right]_{\ell r}p^2\mathcal{P}_{\mu\nu}^T
   +\delta_{\ell r}\frac{1}{\xi}p_\mu p_\nu
   \right\},
\notag\\
   &\left\langle A_\mu^{mn}(x)A_\nu^{pq}(y)\right\rangle
\notag\\
   &=\frac{\lambda}{N}\delta_{mq}\delta_{np}
   \int\frac{d^3p}{(2\pi)^3}
   \frac{1}{2\pi R}\sum_{p_3}
\notag\\
   &\qquad\qquad{}
   \times\left\{e^{ip(x-y)}
   \frac{1}{(p^2)^2}
   \left[
   (1-L)^{-1}p^2\mathcal{P}_{\mu\nu}^L
   +(1-T)^{-1}p^2\mathcal{P}_{\mu\nu}^T
   +\frac{1}{\xi}p_\mu p_\nu
   \right]\right\}_{p\to p_{mn}}.
\label{eq:(2.11)}
\end{align}
Here, we have shown only non-zero propagators. In these expressions, $p_3$
denotes the discrete Kaluza--Klein (KK) momentum along~$S^1$,
\begin{equation}
   p_3=\frac{n}{R},\qquad n\in\mathbb{Z},
\label{eq:(2.12)}
\end{equation}
and the projection operators $\mathcal{P}_{\mu\nu}^T$
and~$\mathcal{P}_{\mu\nu}^L$ are defined by~\cite{Anber:2014sda}
\begin{align}
   &\mathcal{P}_{ij}^T\equiv\delta_{ij}-\frac{p_ip_j}{p^2-p_3^2},\qquad
   \mathcal{P}_{i3}^T=\mathcal{P}_{3i}^T=\mathcal{P}_{33}^T\equiv0,
\notag\\
   &\mathcal{P}_{\mu\nu}^L\equiv\delta_{\mu\nu}-\frac{p_\mu p_\nu}{p^2}
   -\mathcal{P}_{\mu\nu}^T,
\label{eq:(2.13)}
\end{align}
where the Roman letters $i$, $j$, \dots, run only over $0$, $1$, and~$2$, the
uncompactified directions. The functions $L_{\ell r}$, $T_{\ell r}$, $L$,
and~$T$ in~Eq.~\eqref{eq:(2.11)} are given by
\begin{align}
   L_{\ell r}&\equiv
   \frac{\beta_0\lambda}{16\pi^2}
   \Biggl\{
   \delta_{\ell r}\ln\left(\frac{e^{5/3}\mu^2}{p^2}\right)
\notag\\
   &\qquad\qquad\qquad{}
   +12\sum_{j\neq0}(\sigma_{j,N})_{\ell r}\int_0^1dx\,e^{ixp_32\pi Rj}x(1-x)
   \left[K_0(z)-K_2(z)\right]
   \Biggr\},
\notag\\
   T_{\ell r}&\equiv
   \frac{\beta_0\lambda}{16\pi^2}
   \Biggl\{
   \delta_{\ell r}
   \ln\left(\frac{e^{5/3}\mu^2}{p^2}\right)
\notag\\
   &\qquad\qquad\qquad{}
   +12\sum_{j\neq0}(\sigma_{j,N})_{\ell r}\int_0^1dx\,e^{ixp_32\pi Rj}x(1-x)
   \left[K_0(z)-\frac{p_3^2}{p^2}K_2(z)\right]
   \Biggr\},
\notag\\
   L&\equiv
   \frac{\beta_0\lambda}{16\pi^2}
   \Biggl\{\ln\left(\frac{e^{5/3}\mu^2}{p^2}\right)
\notag\\
   &\qquad\qquad\qquad{}
   +12\sum_{j\neq0,j=0\bmod N}\int_0^1dx\,e^{ixp_32\pi Rj}x(1-x)
   \left[K_0(z)-K_2(z)\right]
   \Biggr\},
\notag\\
   T&\equiv
   \frac{\beta_0\lambda}{16\pi^2}
   \Biggl\{\ln\left(\frac{e^{5/3}\mu^2}{p^2}\right)
\notag\\
   &\qquad\qquad\qquad{}
   +12\sum_{j\neq0,j=0\bmod N}\int_0^1dx\,e^{ixp_32\pi Rj}x(1-x)
   \left[K_0(z)-\frac{p_3^2}{p^2}K_2(z)\right]
   \Biggr\},
\label{eq:(2.14)}
\end{align}
where $K_\nu(z)$ denotes the modified Bessel function of the second kind and
the variable~$z$ is defined by
\begin{equation}
   z\equiv\sqrt{x(1-x)}\sqrt{p^2R^2}2\pi|j|.
\label{eq:(2.15)}
\end{equation}
In the first two expressions in~Eq.~\eqref{eq:(2.14)}, $\sigma_{j,N}$ are
$(N-1)\times(N-1)$ real symmetric matrices whose components are defined
by~\cite{Ashie:2019cmy}
\begin{align}
   &(\sigma_{j,N})_{\ell r}
   \equiv
   \frac{1}{N}
   \sum_{m,n=1}^N(\nu^m-\nu^n)_\ell(\nu^m-\nu^n)_r
   e^{i(n-m)2\pi j/N}
\notag\\
   &=\begin{cases}
   \delta_{\ell r},\qquad\text{for $j=0\bmod N$},\\
   -\frac{1}{N}\frac{1}{\sqrt{\ell(\ell+1)r(r+1)}}\Real\left[
   \left(\dfrac{e^{-i\ell2\pi j/N}-1}{e^{-i2\pi j/N}-1}
   -\ell e^{-i\ell2\pi j/N}\right)
   \left(\dfrac{e^{ir2\pi j/N}-1}{e^{i2\pi j/N}-1}
   -r e^{ir2\pi j/N}\right)\right],\\
   \qquad\qquad\qquad\qquad\qquad\qquad\qquad\qquad\qquad\qquad\qquad\qquad
   \text{for $j\neq0\bmod N$}.
   \end{cases}
\label{eq:(2.16)}
\end{align}
In this expression, $\nu^m$ is the $SU(N)$ weights, i.e., the diagonal elements
of Cartan generators $(\nu^m)_\ell\equiv(H_\ell)_{mm}$ (no sum over~$m$ is taken
here). With the convention in~Ref.~\cite{Ashie:2019cmy} (which we adopt
throughout this paper), we have the relations
\begin{equation}
   \sum_{\ell=1}^{N-1}(\nu^m)_\ell(\nu^n)_\ell=\frac{1}{2}\delta_{mn}-\frac{1}{2N}
\label{eq:(2.17)}
\end{equation}
and
\begin{equation}
   \sum_{m=1}^N(\nu^m)_\ell(\nu^m)_r=\frac{1}{2}\delta_{\ell r},\qquad
   \sum_{m=1}^N(\nu^m)_\ell=0.
\label{eq:(2.18)}
\end{equation}
In the W-boson propagator (the second expression in~Eq.~\eqref{eq:(2.11)}), the
momentum variable~$p$ inside the curly brackets is replaced by the twisted
momentum,
\begin{equation}
   p_{mn,\mu}\equiv p_\mu-\delta_{\mu3}\frac{m-n}{RN},\qquad m\neq n,
\label{eq:(2.19)}
\end{equation}
as a consequence of the twisted boundary condition.

In~Eq.~\eqref{eq:(2.14)}, the terms containing the Bessel functions correspond
to  the modifications due to the $S^1$~compactification. If we simply discard
these terms, then $L_{\ell r}=T_{\ell r}$ and~$L=T$, and Eq.~\eqref{eq:(2.11)}
reduces to~Eq.~\eqref{eq:(2.2)} from~Eq.~\eqref{eq:(2.13)} (under the
prescription that $1/(2\pi R)\sum_{p_3}\to\int\frac{dp_3}{2\pi}$). As we have
already noted, for the existence of the IR renormalon, the logarithmic
factor~$\ln p^2$ in the vacuum polarization around~$p^2=0$ is crucial. Here, we
note that $p^2$ can be zero in the vacuum polarization of the photon (the first
expression in~Eq.~\eqref{eq:(2.11)}), whereas it cannot be zero in that of the
W-boson (the second expression in~Eq.~\eqref{eq:(2.11)}). This is because the
momentum of the W-boson is replaced by~Eq.~\eqref{eq:(2.19)} and $p_{mn,3}$
cannot vanish for finite~$RN$. Hence, the W-boson vacuum polarization does not
give rise to an IR renormalon. From these considerations, it is natural to ask
how the logarithmic factor~$\ln p^2$ in the photon vacuum polarization, which
exists in the uncompactified spacetime $\mathbb{R}^4$, is affected by the
$S^1$~compactification~\cite{Anber:2014sda}. This question was studied
in~Ref.~\cite{Anber:2014sda}, and it was shown that the logarithmic
factor~$\ln p^2$ disappears by the effect of the $S^1$~compactification (with
a somewhat different treatment of the gauge field loops to ours) and that there
is no IR renormalon; this was shown for~$N=2$ and~$3$. In the next section, we
explicitly generalize this result of~Ref.~\cite{Anber:2014sda} to arbitrary
finite~$N$. We also comment on how the statement in~Ref.~\cite{Ashie:2019cmy}
that the IR renormalon exists in the $N=\infty$ system
in~$\mathbb{R}^3\times S^1$ should be understood.

\section{Asymptotic behavior of the photon vacuum polarization
in~$\mathbb{R}^3\times S^1$}
\label{sec:3}
In this section, we show that the logarithmic factor~$\ln p^2$ at~$p^2\to0$
disappears in the vacuum polarization of the photon, given by the first two
expressions in~Eq.~\eqref{eq:(2.14)}, for arbitrary finite~$N$. Since $p^2=0$
can be realized only when $p_3=0$, we exclusively assume $p_3=0$ in the
following.

\subsection{Properties of~$\sigma_{j,N}$}
\label{sec:3.1}
To investigate the finite volume effect parts (terms containing the modified
Bessel functions in Eq.~\eqref{eq:(2.14)}), we first study the properties of
the matrix~$\sigma_{j,N}$ defined by~Eq.~\eqref{eq:(2.16)}. From the
definition~\eqref{eq:(2.16)}, $\sigma_{j,N}$ is periodic in~$j$ with the
period~$N$, i.e.,
\begin{equation}
   \sigma_{j+N,N}=\sigma_{j,N}.
\label{eq:(3.1)}
\end{equation}
Note also that
\begin{equation}
   \sigma_{N-j,N}=\sigma_{j,N}.
\label{eq:(3.2)}
\end{equation}

From~Eq.~\eqref{eq:(2.16)}, we also have
\begin{equation}
   \sigma_{j,N}=\mathbbm{1},\qquad\text{for $j=0\bmod N$},
\label{eq:(3.3)}
\end{equation}
where $\mathbbm{1}$ denotes the unit matrix.

For the sum over~$j$, we have
\begin{equation}
   \sum_{j=1}^N\sigma_{j,N}=0
\label{eq:(3.4)}
\end{equation}
and
\begin{equation}
   \sum_{j=1}^Nj\sigma_{j,N}=\frac{N}{2}\mathbbm{1}.
\label{eq:(3.5)}
\end{equation}
Equation~\eqref{eq:(3.4)} immediately follows from the identity
\begin{equation}
   \sum_{j=1}^Ne^{i(n-m)2\pi j/N}=N\delta_{n,m}
\label{eq:(3.6)}
\end{equation}
and the definition~\eqref{eq:(2.16)}, because $n=m$ terms do not contribute
in~Eq.~\eqref{eq:(2.16)}. To see~Eq.~\eqref{eq:(3.5)}, we note
\begin{equation}
   \sum_{j=1}^Nj\frac{1}{2}
   \left[e^{i(n-m)2\pi j/N}+e^{-i(n-m)2\pi j/N}\right]
   =\begin{cases}
   \frac{1}{2}N(N+1),&\text{for $n=m$},\\
   \frac{1}{2}N,&\text{for $n\neq m$},\\
   \end{cases}
\label{eq:(3.7)}
\end{equation}
and thus from the definition~\eqref{eq:(2.16)},
\begin{align}
   \sum_{j=1}^Nj(\sigma_{j,N})_{\ell r}
   &=\frac{1}{2}\sum_{m,n=1}^N
   (\nu^m-\nu^n)_\ell(\nu^m-\nu^n)_r
\notag\\
   &=\frac{N}{2}\delta_{\ell r},
\label{eq:(3.8)}
\end{align}
where we have used~Eq.~\eqref{eq:(2.18)}.

Another interesting property of~$\sigma_{j,N}$ is that they commute to each
other:
\begin{equation}
   [\sigma_{j,N},\sigma_{k,N}]=0.
\label{eq:(3.9)}
\end{equation}
Therefore, all $\sigma_{j,N}$ ($j=0$, $1$, \dots) can be diagonalized by making
use of an orthogonal transformation on the gauge potential~$A_\mu^\ell$
in~Eq.~\eqref{eq:(2.10)}. Equation~\eqref{eq:(3.9)} is obvious
when~$j=0\bmod N$ and/or $k=0\bmod N$, because of~Eq.~\eqref{eq:(3.3)}. For
$j=1$, \dots, $N-1$ and $k=1$, \dots, $N-1$, Eq.~\eqref{eq:(3.9)} can be seen
from the fact that the matrix product
\begin{equation}
   (\sigma_{j,N}\sigma_{k,N})_{\ell r}
   =\begin{cases}
   0,&\text{when $j+k\neq0\bmod N$},\\
   \frac{1}{N^2}
   \sum_{m,n}(\nu^m)_\ell(\nu^n)_r\cos[(m-n)2\pi j/N],
   &\text{when $j+k=0\bmod N$},\\
   \end{cases}
\label{eq:(3.10)}
\end{equation}
which follows from~Eq.~\eqref{eq:(2.17)}, is symmetric
under~$j\leftrightarrow k$.

\subsection{Asymptotic behavior of the photon vacuum polarization
for~$p^2R^2\ll1$}
\label{sec:3.2}
We now study the asymptotic behaviors of the functions $L_{\ell r}$
and~$T_{\ell r}$, which are contained in the photon vacuum polarization. Here,
we introduce the function
\begin{equation}
   f_\nu(p^2R^2)_{\ell r}
   \equiv24\sum_{j=1}^\infty(\sigma_{j,N})_{\ell r}
   \int_0^1dx\,x(1-x)
   K_\nu(\sqrt{x(1-x)}\sqrt{p^2R^2}2\pi j),
\label{eq:(3.11)}
\end{equation}
with $\nu=0$ or~$2$, which corresponds to finite volume corrections. Then, the
functions $L_{\ell r}$ and~$T_{\ell r}$ in~Eq.~\eqref{eq:(2.14)} with~$p_3=0$ are
represented as
\begin{align}
   L_{\ell r}&=
   \frac{\beta_0\lambda}{16\pi^2}
   \left[
   \delta_{\ell r}\ln\left(\frac{e^{5/3}\mu^2}{p^2}\right)
   +f_0(p^2R^2)_{\ell r}
   -f_2(p^2R^2)_{\ell r}\right],
\notag\\
   T_{\ell r}&=
   \frac{\beta_0\lambda}{16\pi^2}
   \left[
   \delta_{\ell r}\ln\left(\frac{e^{5/3}\mu^2}{p^2}\right)
   +f_0(p^2R^2)_{\ell r}\right].
\label{eq:(3.12)}
\end{align}
We thus study the asymptotic behavior of the
function~$f_\nu(p^2R^2)$~\eqref{eq:(3.11)} with~$\nu=0$ and~$2$
for~$p^2R^2\ll1$.

For this, we insert~$\lim_{\epsilon\to0+}e^{-\epsilon j}=1$
into~Eq.~\eqref{eq:(3.11)}:
\begin{equation}
   f_\nu(p^2R^2)_{\ell r}
   =24\sum_{j=1}^\infty(\sigma_{j,N})_{\ell r}\lim_{\epsilon\to0+}e^{-\epsilon j}
   \int_0^1dx\,x(1-x)
   K_\nu(\sqrt{x(1-x)}\sqrt{p^2R^2}2\pi j).
\label{eq:(3.13)}
\end{equation}
Since $|(\sigma_{j,N})_{\ell r}|$ is bounded (it is periodic in~$j$ with the
period~$N$; see~Eq.~\eqref{eq:(3.1)}) and the modified Bessel function
decreases rapidly $K_\nu(z)\sim e^{-z/2}$, the infinite series
$\sum_{j=1}^\infty(\sigma_{j,N})_{\ell r}e^{-\epsilon j}%
\int_0^1dx\,x(1-x)K_\nu(\sqrt{x(1-x)}\sqrt{p^2R^2}2\pi j)$ converges uniformly
in~$\epsilon\geq0$.\footnote{This can be rigorously proven by an argument
similar to that in~Appendix~B of~Ref.~\cite{Ashie:2019cmy}.} This allows us to
exchange the infinite sum~$\sum_{j=1}^\infty$ and the limit~$\lim_{\epsilon\to0+}$
as
\begin{equation}
   f_\nu(p^2R^2)_{\ell r}
   =\lim_{\epsilon\to0+}24\sum_{j=1}^\infty(\sigma_{j,N})_{\ell r}e^{-\epsilon j}
   \int_0^1dx\,x(1-x)
   K_\nu(\sqrt{x(1-x)}\sqrt{p^2R^2}2\pi j).
\label{eq:(3.14)}
\end{equation}

Next, we use the series expansion of the modified Bessel function
\begin{equation}
   K_\nu(z)
   =\sum_{k=0}^\infty\left[b_k^{(\nu)}+c_k^{(\nu)}\ln z\right]z^{2k+\nu}
   +\sum_{k=0}^{\nu-1}d_k^{(\nu)}z^{2k-\nu},
\label{eq:(3.15)}
\end{equation}
in~Eq.~\eqref{eq:(3.14)} (for $\nu=0$, the second sum in~Eq.~\eqref{eq:(3.15)}
is set to zero). Here, the first few coefficients are given by
\begin{equation}
   b_0^{(0)}=\ln2-\gamma,\qquad
   c_0^{(0)}=-1,
\label{eq:(3.16)}
\end{equation}
where $\gamma$ is the Euler--Mascheroni constant, and
\begin{equation}
   d_0^{(2)}=2,\qquad
   d_1^{(2)}=-\frac{1}{2}.
\label{eq:(3.17)}
\end{equation}
Note that with the substitution~$z=\sqrt{x(1-x)}\sqrt{p^2R^2}2\pi j$,
Eq.~\eqref{eq:(3.15)} becomes the series expansion in~$p^2R^2$
(and~$\log(p^2 R^2)$). Then, since the damping factor~$e^{-\epsilon j}$ provides
a good convergence property for the $j$-summation, we intuitively expect that
the $j$-summation can be done for each term in the $k$-summation. This naive
exchange of the $j$-summation and the $k$-summation yields
\begin{align}
   f_0(p^2R^2)_{\ell r}
   &=2\lim_{\epsilon\to0+}\sum_{j=1}^\infty(\sigma_{j,N})_{\ell r}e^{-\epsilon j}
   \left[-\ln(p^2R^2)
   +\frac{5}{3}-2\ln\pi-2\gamma
   -2\ln j\right]
\notag\\
   &\qquad{}+O(p^2R^2\ln(p^2R^2)),
\notag\\
   f_2(p^2R^2)_{\ell r}
   &=2\lim_{\epsilon\to0+}\sum_{j=1}^\infty(\sigma_{j,N})_{\ell r}e^{-\epsilon j}
   \left(
   \frac{6}{\pi^2j^2}\frac{1}{p^2R^2}
   -1\right)+O(p^2R^2\ln(p^2R^2)).
\label{eq:(3.18)}
\end{align}

In fact, it is not easy to give a rigorous justification for the above exchange
of the $j$- and $k$-summations (even in the sense of the asymptotic expansion)
or, in other words, to show that the last remaining terms
in~Eq.~\eqref{eq:(3.18)} are really $O(p^2R^2\ln(p^2R^2))$. In
Appendix~\ref{sec:A}, we give a rigorous proof for the leading asymptotic
behaviors of~$f_{\nu}(p^2 R^2)_{\ell r}$ with $\nu=0$ and~$2$ for~$p^2R^2\ll1$ up
to~$O((p^2R^2)^0)$ terms; the results are indeed consistent
with~Eq.~\eqref{eq:(3.18)} and consequently also with
Eqs.~\eqref{eq:(3.29)}--\eqref{eq:(3.31)} below. This proof is sufficient to
conclude the disappearance of the logarithmic factor~$\ln p^2$ in the photon
vacuum polarization. For the $O((p^2R^2)^0)$ terms in~Eq.~\eqref{eq:(3.18)}, we
do not have a rigorous proof, although it is highly plausible that the above
exchange of the $j$- and $k$-summations is legitimate; we also numerically
check that Eqs.~\eqref{eq:(3.29)}--\eqref{eq:(3.31)} are indeed correct for
some small~$N$.

Now, in the first expression in~Eq.~\eqref{eq:(3.18)}, the sum of terms not
containing~$\ln j$ can be computed as
\begin{align}
   \sum_{j=1}^\infty(\sigma_{j,N})_{\ell r}e^{-\epsilon j}
   &=\sum_{j=1}^N(\sigma_{j,N})_{\ell r}e^{-\epsilon j}
   \sum_{b=0}^\infty e^{-\epsilon bN}
\notag\\
   &=\sum_{j=1}^{N}(\sigma_{j,N})_{\ell r}e^{-\epsilon j}
   \frac{1}{1-e^{-\epsilon N}}
\notag\\
   &=\sum_{j=1}^{N}(\sigma_{j,N})_{\ell r}
   \left[
   \frac{1}{N\epsilon}+\frac{1}{2}-\frac{j}{N}
   +O(\epsilon)
   \right]
\notag\\
   &=-\frac{1}{2}\delta_{\ell r}+O(\epsilon),
\label{eq:(3.19)}
\end{align}
where in the first equality we have used the fact that $\sigma_{j,N}$ is
periodic in~$j$ with the period~$N$; in the last step, we have used the
properties~\eqref{eq:(3.4)} and~\eqref{eq:(3.5)}. Therefore, we obtain
\begin{equation}
   \lim_{\epsilon\to0+}\sum_{j=1}^\infty(\sigma_{j,N})_{\ell r}e^{-\epsilon j}
   =-\frac{1}{2}\delta_{\ell r}.
\label{eq:(3.20)}
\end{equation}

On the other hand, the sum over terms containing~$\ln j$ in the first equation
of~Eq.~\eqref{eq:(3.18)} is evaluated as
\begin{equation}
   \sum_{j=1}^\infty(\sigma_{j,N})_{\ell r}e^{-\epsilon j}\ln j
   =\sum_{j=1}^N(\sigma_{j,N})_{\ell r}e^{-\epsilon j}
   \left[
   \sum_{b=1}^\infty e^{-\epsilon bN}\ln(j+bN)
   +\ln j\right].
\label{eq:(3.21)}
\end{equation}
The calculation of the first term in the square brackets proceeds as follows:
\begin{align}
   &\sum_{b=1}^\infty
   e^{-\epsilon bN}\ln(j+bN)
\notag\\
   &=\sum_{b=1}^\infty
   e^{-\epsilon bN}
   \left\{
   \left[\ln\left(1+\frac{j}{bN}\right)-\frac{j}{bN}\right]
   +\frac{j}{bN}+\ln(bN)
   \right\}
\notag\\
   &=-\gamma\frac{j}{N}-\ln(j/N)-\ln{\mit\Gamma}(j/N)
   -\frac{j}{N}\ln(\epsilon N)+O(\epsilon)
   +\sum_{b=1}^\infty
   e^{-\epsilon bN}\ln(bN).
\label{eq:(3.22)}
\end{align}
We compute the last infinite sum as
\begin{align}
   \sum_{b=1}^\infty
   e^{-\epsilon bN}\ln(bN)
   &=-\frac{\partial}{\partial s}
   \left[\sum_{b=1}^\infty e^{-\epsilon bN}(bN)^{-s}\right]_{s=0}
\notag\\
   &=-\frac{\partial}{\partial s}
   \left[N^{-s}\Li_s(e^{-\epsilon N})\right]_{s=0},
\label{eq:(3.23)}
\end{align}
where $\Li_s(z)$ is the polylogarithm function
\begin{equation}
   \Li_s(z)\equiv\sum_{n=1}^\infty\frac{z^n}{n^s},\qquad\text{for $|z|<1$}.
\label{eq:(3.24)}
\end{equation}
Then, using the expansion
\begin{equation}
   \Li_s(z)
   ={\mit\Gamma}(1-s)(-\ln z)^{s-1}
   +\sum_{k=0}^\infty\frac{\zeta(s-k)}{k!}(\ln z)^k,
\label{eq:(3.25)}
\end{equation}
we have
\begin{equation}
   \sum_{b=1}^\infty
   e^{-\epsilon bN}\ln(bN)
   =-\frac{1}{\epsilon N}(\ln\epsilon+\gamma)
   -\frac{1}{2}\ln\frac{N}{2\pi}
   +O(\epsilon).
\label{eq:(3.26)}
\end{equation}
Using this in~Eq.~\eqref{eq:(3.22)}, Eq.~\eqref{eq:(3.21)} is given by
\begin{align}
   &\sum_{j=1}^\infty(\sigma_{j,N})_{\ell r}e^{-\epsilon j}\ln j
\notag\\
   &=\sum_{j=1}^N(\sigma_{j,N})_{\ell r}
   \left[
   -\frac{j}{N}\ln N-\ln{\mit\Gamma(j/N)}
   -\frac{1}{\epsilon N}(\ln\epsilon+\gamma)
   +\ln\sqrt{2\pi N}
   +O(\epsilon)\right]
\notag\\
   &=-\frac{1}{2}\ln N\delta_{\ell r}
   -\sum_{j=1}^N(\sigma_{j,N})_{\ell r}\ln{\mit\Gamma(j/N)}+O(\epsilon),
\label{eq:(3.27)}
\end{align}
where we have used~Eqs.~\eqref{eq:(3.4)} and~\eqref{eq:(3.5)}. In this way,
we obtain
\begin{equation}
   \lim_{\epsilon\to0+}\sum_{j=1}^\infty(\sigma_{j,N})_{\ell r}e^{-\epsilon j}\ln j
   =-\frac{1}{2}\ln N\delta_{\ell r}
   -\sum_{j=1}^{N-1}(\sigma_{j,N})_{\ell r}\ln{\mit\Gamma}(j/N).
\label{eq:(3.28)}
\end{equation}

Finally, by combining Eqs.~\eqref{eq:(3.18)}, \eqref{eq:(3.20)},
and~\eqref{eq:(3.28)}, we obtain the asymptotic form for~$p^2R^2\ll1$,
\begin{align}
   &f_0(p^2R^2)_{\ell r}
\notag\\
   &=\left[\ln(p^2R^2)-\frac{5}{3}+2\ln\pi+2\gamma+2\ln N\right]\delta_{\ell r}
   +4\sum_{j=1}^{N-1}(\sigma_{j,N})_{\ell r}\ln{\mit\Gamma}(j/N)
   +O(p^2R^2\ln(p^2R^2)),
\label{eq:(3.29)}
\end{align}
i.e.,
\begin{align}
   &\delta_{\ell r}\ln\left(\frac{e^{5/3}}{p^2R^2}\right)
   +f_0(p^2R^2)_{\ell r}
\notag\\
   &=(2\gamma+2\ln N+2\ln\pi)\delta_{\ell r}
   +4\sum_{j=1}^{N-1}(\sigma_{j,N})_{\ell r}\ln{\mit\Gamma}(j/N)
   +O(p^2R^2\ln(p^2R^2)),
\label{eq:(3.30)}
\end{align}
and in a similar way
\begin{equation}
   f_2(p^2R^2)_{\ell r}
   =\frac{12}{\pi^2}\sum_{j=1}^\infty(\sigma_{j,N})_{\ell r}\frac{1}{j^2}
   \frac{1}{p^2R^2}
   +\delta_{\ell r}
   +O(p^2R^2\ln(p^2R^2)).
\label{eq:(3.31)}
\end{equation}
These are our main results in the first part of this paper. Since the photon
vacuum polarization with~$p_3=0$ is given by
\begin{align}
   L_{\ell r}&=
   \frac{\beta_0\lambda}{16\pi^2}
   \left[
   \delta_{\ell r}\ln(\mu^2R^2)
   +\delta_{\ell r}\ln\left(\frac{e^{5/3}}{p^2R^2}\right)
   +f_0(p^2R^2)_{\ell r}
   -f_2(p^2R^2)_{\ell r}\right],
\notag\\
   T_{\ell r}&=
   \frac{\beta_0\lambda}{16\pi^2}
   \left[
   \delta_{\ell r}\ln(\mu^2R^2)
   +\delta_{\ell r}\ln\left(\frac{e^{5/3}}{p^2R^2}\right)
   +f_0(p^2R^2)_{\ell r}\right],
\label{eq:(3.32)}
\end{align}
it is shown with~Eqs.~\eqref{eq:(3.30)} and~\eqref{eq:(3.31)} that the
logarithmic factor~$\ln p^2$ disappears in the photon vacuum polarization
because of the effect of the $S^1$~compactification.\footnote{The $1/(p^2R^2)$
behavior in~$L_{\ell r}$ implies that naive perturbation theory suffers from IR
divergences. To see this, one should note that $L_{\ell r}$ is~$O(\lambda)$
and thus higher powers of~$L_{\ell r}$ are included in the numerator of the
integrand in calculating higher-order perturbative coefficients; see the first
equation of~Eq.~\eqref{eq:(2.11)}. Hence, we have severer IR divergences at
higher orders. This is nothing but the famous IR divergence at finite
temperature~\cite{Gross:1980br,Kapusta:2006pm} although here the boundary
condition for the adjoint fermion is not anti-periodic. To avoid the IR
divergences, the $1/(p^2R^2)$ term should not be expanded and should be kept in
the denominator. With this understanding, the absence of the IR renormalon is
concluded. See also Sect.~\ref{sec:4}.\label{footnote:6}} This generalizes the
observation made in~Ref.~\cite{Anber:2014sda} for~$N=2$ and~$3$ to arbitrary
finite~$N$. Hence, it is concluded that the IR renormalon is absent for
arbitrary finite~$N$. (A supplementary explanation for the absence of the IR
renormalon is given in Sect.~\ref{sec:4} with an explicit example of a
gauge-invariant quantity.)

In the subsequent subsections, we will show the asymptotic behaviors explicitly
for some small $N$.

\subsection{$N=2$}
\label{sec:3.3}
For~$N=2$, $\ell$ and~$r$ can take only $\ell=r=1$ in~Eq.~\eqref{eq:(2.16)} and
\begin{equation}
   (\sigma_{j,2})_{11}
   =\begin{cases}
   1,&\text{for $j=0\bmod2$,}\\
   -1,&\text{for $j=1\bmod2$.}\\
   \end{cases}
\label{eq:(3.33)}
\end{equation}
From this, we have
\begin{align}
   \sum_{j=1}^1(\sigma_{j,2})_{11}\ln{\mit\Gamma}(j/2)
   &=-\ln\sqrt{\pi},
\notag\\
   \sum_{j=1}^\infty(\sigma_{j,2})_{11}\frac{1}{j^2}
   &=-\frac{\pi^2}{12},
\label{eq:(3.34)}
\end{align}
and, therefore, from~Eqs.~\eqref{eq:(3.30)} and~\eqref{eq:(3.31)},
\begin{align}
   \ln\left(\frac{e^{5/3}}{p^2R^2}\right)+f_0(p^2R^2)_{11}
   &=2\gamma+\ln4+O(p^2R^2\ln(p^2R^2)),
\notag\\
   f_2(p^2R^2)_{11}
   &=-\frac{1}{p^2R^2}+1+O(p^2R^2\ln(p^2R^2)).
\label{eq:(3.35)}
\end{align}

\begin{figure}[htbp]
\centering
\includegraphics[width=0.7\columnwidth]{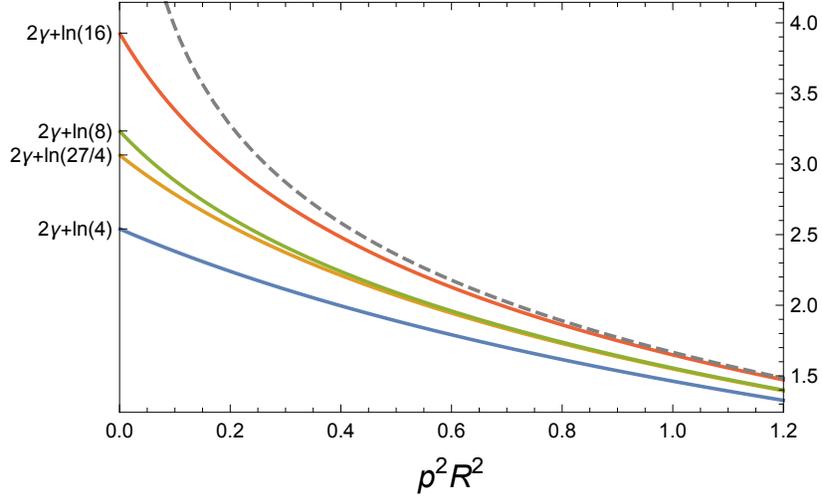}
\caption{The function $\delta_{\ell r}\ln[e^{5/3}/(p^2R^2)]+f_0(p^2R^2)_{\ell r}$.
From bottom to top,
$\ln[e^{5/3}/(p^2R^2)]+f_0(p^2R^2)_{11}$ for~$N=2$,
$\ln[e^{5/3}/(p^2R^2)]+f_0(p^2R^2)_{11}$ for~$N=3$, and
$\ln[e^{5/3}/(p^2R^2)]+f_0(p^2R^2)_{11}$ and
$\ln[e^{5/3}/(p^2R^2)]+f_0(p^2R^2)_{33}$ for~$N=4$.
The logarithmic function~$\ln[e^{5/3}/(p^2R^2)]$ is also shown by the broken
line.}
\label{fig:1}
\end{figure}

\begin{figure}[htbp]
\centering
\includegraphics[width=0.7\columnwidth]{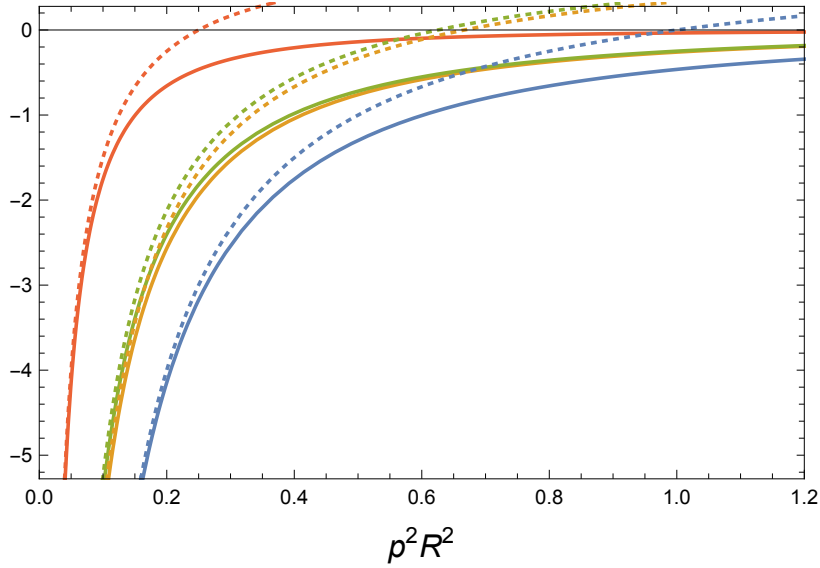}
\caption{The function $f_2(p^2R^2)_{\ell r}$.
From bottom to top,
$f_2(p^2R^2)_{11}$ for~$N=2$,
$f_2(p^2R^2)_{11}$ for~$N=3$, and
$f_2(p^2R^2)_{11}$ and
$f_2(p^2R^2)_{33}$ for~$N=4$.
The dashed lines show their asymptotic behaviors 
obtained from~Eqs.~\eqref{eq:(3.35)}, \eqref{eq:(3.38)}, and \eqref{eq:(3.44)}.}
\label{fig:2}
\end{figure}

In Figs.~\ref{fig:1} and~\ref{fig:2}, we plot the functions appearing on the
left-hand side of~Eq.~\eqref{eq:(3.35)}. We numerically compute them directly
from the definition~\eqref{eq:(3.11)}. In~Fig.\ref{fig:1}, we see that the
asymptotic value of~$\ln[e^{5/3}/(p^2R^2)]+f_0(p^2R^2)_{11}$ (the blue curve)
as~$p^2R^2\to0$ is correctly given by~Eq.~\eqref{eq:(3.35)}. The broken line
shows the logarithmic function~$\ln[e^{5/3}/(p^2R^2)]$, which diverges
as~$p^2R^2\to0$. In~Fig.~\ref{fig:2}, we show the function~$f_2(p^2R^2)_{11}$ by
a solid line (blue), and we can see that it indeed approaches the asymptotic
behavior obtained in~Eq.~\eqref{eq:(3.35)}, which is shown by the dashed line
of the same color, as~$p^2R^2\to0$.

\subsection{$N=3$}
\label{sec:3.4}
In this case with~$N=3$, from~Eq.~\eqref{eq:(2.16)}, we have
\begin{align}
   &(\sigma_{j,3})_{\ell r}=
   \delta_{\ell r}\begin{cases}
   1,&\text{for $j=0\bmod3$},\\
   -\frac{1}{2},&\text{for $j=1$, $2\bmod3$},\\
   \end{cases}
\label{eq:(3.36)}
\end{align}
and then
\begin{align}
   \sum_{j=1}^2(\sigma_{j,3})_{\ell r}\ln{\mit\Gamma}(j/3)
   &=-\frac{1}{2}\ln\left(\frac{2\pi}{\sqrt{3}}\right)
   \delta_{\ell r},
\notag\\
   \sum_{j=1}^\infty(\sigma_{j,3})_{\ell r}\frac{1}{j^2}
   &=-\frac{\pi^2}{18}\delta_{\ell r}.
\label{eq:(3.37)}
\end{align}
We thus obtain
\begin{align}
   \delta_{\ell r}\ln\left(\frac{e^{5/3}}{p^2R^2}\right)+f_0(p^2R^2)_{\ell r}
   &=\left[2\gamma+\ln\frac{27}{4}+O(p^2R^2\ln(p^2R^2))
   \right]\delta_{\ell r},
\notag\\
   f_2(p^2R^2)_{\ell r}
   &=\left[-\frac{2}{3}\frac{1}{p^2R^2}+1+O(p^2R^2\ln(p^2R^2))
   \right]\delta_{\ell r}.
\label{eq:(3.38)}
\end{align}

In Figs.~\ref{fig:1} and~\ref{fig:2}, we plot the functions appearing on the
left-hand side of~Eq.~\eqref{eq:(3.38)}. Again, in~Fig.~\ref{fig:1}, we see
that the asymptotic value of~$\ln[e^{5/3}/(p^2R^2)]+f_0(p^2R^2)_{11}$ (the yellow
curve) as~$p^2R^2\to0$ is correctly given by~Eq.~\eqref{eq:(3.38)}. Also
in~Fig.~\ref{fig:2}, we can confirm the validity of the asymptotic form
of~$f_2(p^2 R^2)_{11}$.

We can compare the results in~Ref.~\cite{Anber:2014sda} for $N=2$ and~$3$ with
our results in~Eqs.~\eqref{eq:(3.35)} and~\eqref{eq:(3.38)}. Using
$\beta_0=11/3-2n_W/3$, Eqs.~(5.3) and~(5.4) of~Ref.~\cite{Anber:2014sda} show
that for~$p_3=0$ (in our notation)
\begin{align}
   L_{\ell r}
   &=\frac{(\beta_0-3)\lambda}{16\pi^2}
   \left\{\left[1-3\left(\frac{2}{N}-1\right)^2\right]\frac{1}{p^2R^2}
   -\frac{2}{3}\right\}\delta_{\ell r}
\notag\\
   &\qquad{}
   +\frac{\beta_0\lambda}{16\pi^2}
   \left[\ln(\Lambda_0^2R^2)-\ln4-\psi(1/N)-\psi(1-1/N)\right]\delta_{\ell r}
   +O(p^2R^2\ln(p^2R^2)),
\notag\\
   T_{\ell r}
   &=\frac{(\beta_0-3)\lambda}{16\pi^2}
   \frac{1}{3}\delta_{\ell r}
   +\frac{\beta_0\lambda}{16\pi^2}
   \left[\ln(\Lambda_0^2R^2)-\ln4-\psi(1/N)-\psi(1-1/N)\right]\delta_{\ell r}
\notag\\
   &\qquad{}
   +O(p^2R^2\ln(p^2R^2)),
\label{eq:(3.39)}
\end{align}
where $\Lambda_0$ is the renormalization scale in~Ref.~\cite{Anber:2014sda}
and~$\psi(z)\equiv(d/dz)\ln{\mit\Gamma}(z)$. Noting that
$\psi(1/2)=-\gamma-\ln4$ and~$\psi(1/3)+\psi(2/3)=-2\gamma-3\ln3$, we see that
by choosing
\begin{equation}
   \Lambda_0^2=e^{-1/3}\mu^2,
\label{eq:(3.40)}
\end{equation}
these expressions perfectly coincide with our results, Eq.~\eqref{eq:(3.32)}
with~Eqs.~\eqref{eq:(3.35)} and~\eqref{eq:(3.38)}, for~$\beta_0\to\infty$. We
note that, because of the difference in the treatment of the gauge field loops,
we expect that the results in~Ref.~\cite{Anber:2014sda} coincide with ours only
in the limit~$\beta_0\to\infty$, in which the contribution of the fermion loop
diagrams dominates.

\subsection{$N=4$}
\label{sec:3.5}
For this case, Eq.~\eqref{eq:(2.16)} gives
\begin{equation}
   (\sigma_{j,4})_{\ell r}
   =\begin{cases}
   \begin{pmatrix}
   1&0&0
\\
   0&1&0
\\
   0&0&1
\\
   \end{pmatrix},
   &\text{for $j=0\bmod4$},
\\
   \begin{pmatrix}
   -\frac{1}{4}&-\frac{1}{4\sqrt{3}}&\frac{1}{2\sqrt{6}}
\\
   -\frac{1}{4\sqrt{3}}&-\frac{5}{12}&-\frac{1}{6\sqrt{2}}
\\
   \frac{1}{2\sqrt{6}}&-\frac{1}{6\sqrt{2}}&-\frac{1}{3}
\\
   \end{pmatrix},
   &\text{for $j=1$, $3\bmod4$},
\\
   \begin{pmatrix}
   -\frac{1}{2}&\frac{1}{2\sqrt{3}}&-\frac{1}{\sqrt{6}}
\\
   \frac{1}{2\sqrt{3}}&-\frac{1}{6}&\frac{1}{3\sqrt{2}}
\\
   -\frac{1}{\sqrt{6}}&\frac{1}{3\sqrt{2}}&-\frac{1}{3}
\\
   \end{pmatrix},
   &\text{for $j=2\bmod4$},\\
   \end{cases},
\label{eq:(3.41)}
\end{equation}
where the row and the column refer to the indices $\ell$ and~$r$, respectively.
As noted in~Eq.~\eqref{eq:(3.9)}, these matrices can be simultaneously
diagonalized by an orthogonal transformation. After this diagonalization, we
have
\begin{equation}
   (\sigma_{j,4})_{\ell r}
   =\begin{cases}
   \begin{pmatrix}
   1&0&0
\\
   0&1&0
\\
   0&0&1
\\
   \end{pmatrix},
   &\text{for $j=0\bmod4$},
\\
   \begin{pmatrix}
   -\frac{1}{2}&0&0
\\
   0&-\frac{1}{2}&0
\\
   0&0&0
\\
   \end{pmatrix},
   &\text{for $j=1$, $3\bmod4$},
\\
   \begin{pmatrix}
   0&0&0
\\
   0&0&0
\\
   0&0&-1
\\
   \end{pmatrix},
   &\text{for $j=2\bmod4$},\\
   \end{cases}.
\label{eq:(3.42)}
\end{equation}
In this diagonal basis, we have
\begin{align}
   \sum_{j=1}^3(\sigma_{j,4})_{\ell r}\ln{\mit\Gamma}(j/4)
   &=\begin{cases}
   -\frac{1}{2}\ln(\sqrt{2}\pi),&\text{for~$\ell=r=1$},\\
   -\frac{1}{2}\ln(\sqrt{2}\pi),&\text{for~$\ell=r=2$},\\
   -\frac{1}{2}\ln\pi,&\text{for~$\ell=r=3$},\\
   0,&\text{otherwise},\\
   \end{cases}
\notag\\
   \sum_{j=1}^\infty(\sigma_{j,4})_{\ell r}\frac{1}{j^2}
   &=\begin{cases}
   -\frac{5\pi^2}{96},&\text{for~$\ell=r=1$},\\
   -\frac{5\pi^2}{96},&\text{for~$\ell=r=2$},\\
   -\frac{\pi^2}{48},&\text{for~$\ell=r=3$}.\\
   0,&\text{otherwise}.\\
   \end{cases}
\label{eq:(3.43)}
\end{align}
We thus have
\begin{align}
\notag\\
   \delta_{\ell r}\ln\left(\frac{e^{5/3}}{p^2R^2}\right)+f_0(p^2R^2)_{\ell r}
   &=
   \begin{cases}
   2\gamma+\ln8+O(p^2R^2\ln(p^2R^2)),&\text{for~$\ell=r=1$},\\
   2\gamma+\ln8+O(p^2R^2\ln(p^2R^2)),&\text{for~$\ell=r=2$},\\
   2\gamma+\ln16+O(p^2R^2\ln(p^2R^2)),&\text{for~$\ell=r=3$},\\
   0,&\text{otherwise}.\\
   \end{cases}
\notag\\
   f_2(p^2R^2)_{\ell r}
   &=
   \begin{cases}
   -\frac{5}{8}\frac{1}{p^2R^2}
   +1+O(p^2R^2\ln(p^2R^2)),&\text{for~$\ell=r=1$},\\
   -\frac{5}{8}\frac{1}{p^2R^2}
   +1+O(p^2R^2\ln(p^2R^2)),&\text{for~$\ell=r=2$},\\
   -\frac{1}{4}\frac{1}{p^2R^2}
   +1+O(p^2R^2\ln(p^2R^2)),&\text{for~$\ell=r=3$},\\
   0,&\text{otherwise}.\\
   \end{cases}
\label{eq:(3.44)}
\end{align}

In Figs.~\ref{fig:1} and~\ref{fig:2}, we plot the functions appearing on the
left-hand side of~Eq.~\eqref{eq:(3.44)}. In~Fig.\ref{fig:1}, we see that the
asymptotic value of~$\delta_{\ell r}\ln[e^{5/3}/(p^2R^2)]+f_0(p^2R^2)_{\ell r}$
(the green and orange curves) as~$p^2R^2\to0$ is correctly given
by~Eq.~\eqref{eq:(3.44)}. Also in~Fig.~\ref{fig:2}, we can confirm the validity
of the asymptotic form of~$f_2(p^2 R^2)_{11}$ and~$f_2(p^2 R^2)_{33}$.

\subsection{Comment on the $N=\infty$ case}
\label{sec:3.6}
In~Ref.~\cite{Ashie:2019cmy}, the $N\to\infty$ limit of the expressions
in~Eq.~\eqref{eq:(2.14)} is considered and it is concluded that the IR
renormalon exists in this limit. Since we have observed that there is no IR
renormalon for arbitrary finite~$N$, we should clarify how these two
conclusions are related. The crucial relation that led to the existence of the
IR renormalon for~$N\to\infty$ is the bounds~\cite{Ashie:2019cmy}
\begin{equation}
   \left|\sum_{j\neq0}
   \sigma_{j,N}\int_0^1dx\,e^{ixp_32\pi Rj}x(1-x)K_0(z)\right|
   <\frac{8\zeta(3)}{\pi^3(p^2R^2)^{3/2}}
   \left(\frac{1}{N^3}+\frac{4}{N}\right)
\label{eq:(3.45)}
\end{equation}
and
\begin{align}
   &\left|\sum_{j\neq0}
   \sigma_{j,N}\int_0^1dx\,e^{ixp_32\pi Rj}x(1-x)K_2(z)\right|
\notag\\
   &<
   \frac{16\zeta(4)}{\pi^4(p^2R^2)^2}
   \left(\frac{1}{N^4}+\frac{4}{N}\right)
   +\frac{8\zeta(3)}{\pi^3(p^2R^2)^{3/2}}
   \left(\frac{1}{N^3}+\frac{4}{N}\right)
\label{eq:(3.46)}.
\end{align}
Similar bounds hold for the finite volume parts in~$L$ and~$T$ for the W-boson
vacuum polarization, i.e., for the expressions where $\sigma_{j,N}$ is omitted
and the sum is replaced by~$\sum_{j\neq0,j=0\bmod N}$. From these bounds, for a
\emph{fixed non-zero\/} momentum~$p$, the terms containing the Bessel functions
in~Eq.~\eqref{eq:(2.14)} (finite volume corrections) vanish as~$N\to\infty$;
the vacuum polarizations then become those in~$\mathbb{R}^4$. This is the basic
logic in~Ref.~\cite{Ashie:2019cmy} in concluding the IR renormalon (see
also~Ref.~\cite{Ishikawa:2019oga}). The problem with this argument is that the
bound is not uniform in the momentum~$p$. The situation can be clearly seen
in~Figs.~\ref{fig:1} and~\ref{fig:2}; for any fixed non-zero~$p^2R^2$, the
functions $f_0(p^2R^2)$ and~$f_2(p^2R^2)$ vanish as~$N\to\infty$. In particular,
in~Fig.~\ref{fig:1}, the curves approach the logarithmic
function~$\ln[e^{5/3}/(p^2R^2)]$ (the broken line) as~$N\to\infty$ at each fixed
non-zero value of~$p^2R^2$. However, in~Fig.~\ref{fig:1}, as long as $N$ is
finite, the limiting value as~$p^2R^2\to0$ is finite and does not have the
logarithmic behavior~$\ln p^2$ at~$p^2=0$. A similar remark applies to the
vacuum polarization of the W-boson, because the logarithmic behavior can appear
only when the twisted shift of the momentum vanishes, i.e., when~$N=\infty$. In
this way, the vacuum polarization may possess the logarithmic factor only at
the limiting point~$N=\infty$; the existence of the IR renormalon is peculiar
in this single point. Therefore, $N=\infty$ cannot be used as a starting point
for the study of IR renormalons in $SU(N)$ QCD(adj.) with finite~$N$ even
though it is very large. In this sense, we have to admit that the statement on
the IR renormalon in~Ref.~\cite{Ashie:2019cmy} is not wrong but misleading.

\section{Decompactification limit~$R\to\infty$ and renormalon precursor}
\label{sec:4}
We have observed that, under the $S^1$~compactification, the
functions~$L_{\ell r}$ and~$T_{\ell r}$ in~Eq.~\eqref{eq:(2.14)} appearing in the
vacuum polarization of the photon with~$p_3=0$ lose the logarithmic
behavior~$\ln p^2$ for~$p^2\to0$. The vacuum polarization of the W-boson also
does not possess $\ln p^2$ behavior around $p^2=0$ because its momentum is
given by the twisted momentum~$p_{mn}$ of~Eq.~\eqref{eq:(2.19)} and $p_{mn}^2$
cannot take zero as long as $RN$ is finite. According to the discussion
in~Sect.~\ref{sec:2}, therefore, there is no ambiguity associated with the IR
renormalon (i.e., the factorial growth of perturbative coefficients) under the
$S^1$~compactification. Then, it is natural to wonder how the ambiguity
associated with the IR renormalon in~$\mathbb{R}^4$ can emerge under the
decompactification of~$S^1$, $R\to\infty$. The purpose of this section is to
understand this issue. We are naturally led to introduce the notion of the
``renormalon precursor'' from this consideration.

\begin{figure}[htbp]
\centering
\begin{subfigure}{0.47\columnwidth}
\centering
\includegraphics[width=\columnwidth]{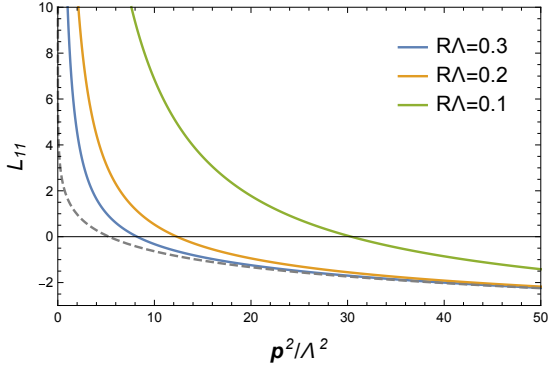}
\caption{$16\pi^2/(\beta_0\lambda)L_{11}(\mu=\Lambda)|_{p_3=0}$
for~$R\Lambda=0.3$, $0.2$, and~$0.1$.}
\label{fig:3a}
\end{subfigure}
\hspace{1em}
\begin{subfigure}{0.47\columnwidth}
\centering
\includegraphics[width=\columnwidth]{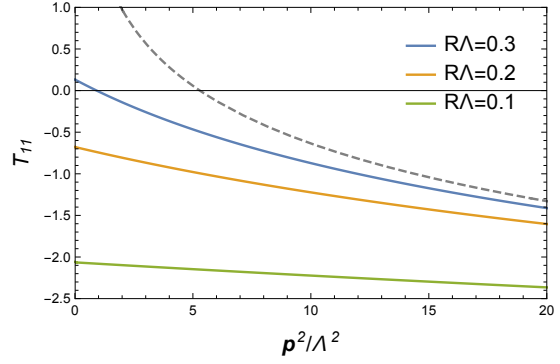}
\caption{$16\pi^2/(\beta_0\lambda)T_{11}(\mu=\Lambda)|_{p_3=0}$
for~$R\Lambda=0.3$, $0.2$, and~$0.1$.}
\label{fig:3b}
\end{subfigure}
\begin{subfigure}{0.47\columnwidth}
\centering
\includegraphics[width=\columnwidth]{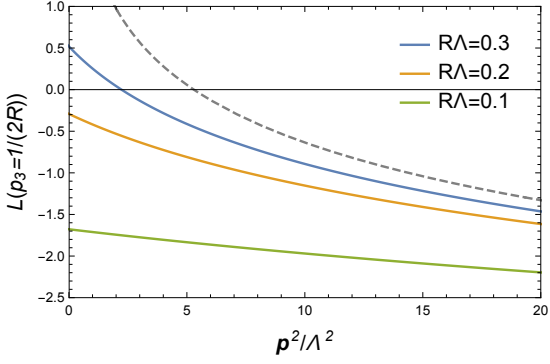}
\caption{$16\pi^2/(\beta_0\lambda)L(\mu=\Lambda)|_{p_3=1/(2R)}$
for~$R\Lambda=0.3$, $0.2$, and~$0.1$.}
\label{fig:3c}
\end{subfigure}
\hspace{1em}
\begin{subfigure}{0.47\columnwidth}
\centering
\includegraphics[width=\columnwidth]{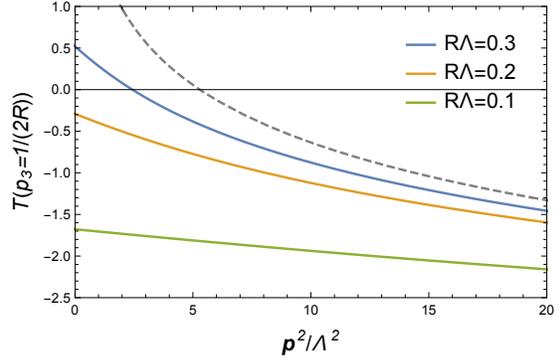}
\caption{$16\pi^2/(\beta_0\lambda)T(\mu=\Lambda)|_{p_3=1/(2R)}$
for~$R\Lambda=0.3$, $0.2$, and~$0.1$.}
\label{fig:3d}
\end{subfigure}
\caption{$(16\pi^2/\beta_0\lambda)L_{11}$, $(16\pi^2/\beta_0\lambda)T_{11}$,
$(16\pi^2/\beta_0\lambda)L$ and $(16\pi^2/\beta_0\lambda)T$ at~$\mu=\Lambda$
(see~Eq.~\eqref{eq:(4.3)}) as functions of~$\bm{p}^2=\sum_{i=0}^2p_i^2$ for
various values of the compactification radius~$R\Lambda$. $p_3$ is set to the
smallest possible values as~Eq.~\eqref{eq:(4.3)}. The dashed lines are the
infinite volume result, $\log(e^{5/3}\Lambda^2/p^2)$.}
\label{fig:3}
\end{figure}

To illustrate the idea of the renormalon precursor, let us consider the example
of the ``gluon condensate'' in the $N=2$ theory that is given in the
large-$\beta_0$ approximation from~Eq.~\eqref{eq:(2.11)} by\footnote{Recall
that for~$N=2$, $\ell$ and~$r$ can take only $\ell=r=1$ in~$L_{\ell r}$
and~$T_{\ell r}$ and these are not matrices but simply numbers.}
\begin{align}
   \left\langle\tr(F_{\mu\nu}F_{\mu\nu})\right\rangle
   &=-\frac{\lambda}{2}
   \int\frac{d^3p}{(2\pi)^3}\frac{1}{2\pi R}\sum_{p_3}
   \left\{
   \left[1-L_{11}(\mu)\right]^{-1}+2\left[1-T_{11}(\mu)\right]^{-1}
   \right\}
\notag\\
   &\qquad{}
   -\frac{\lambda}{2}
   \int\frac{d^3p}{(2\pi)^3}\frac{1}{2\pi R}\sum_{p_3}
   \left\{
   \left[1-L(\mu)\right]^{-1}
   +2\left[1-T(\mu)\right]^{-1}
   \right\}_{p\to p_{12}}
\notag\\
   &=\frac{\lambda}{2}
   \int\frac{d^3p}{(2\pi)^3}\frac{1}{2\pi R}\sum_{p_3}
   \left[
   \frac{1}{L_{11}(\mu=\Lambda)}
   +\frac{2}{T_{11}(\mu=\Lambda)}
   \right]
\notag\\
   &\qquad{}
   +\frac{\lambda}{2}
   \int\frac{d^3p}{(2\pi)^3}\frac{1}{2\pi R}\sum_{p_3}
   \left[
   \frac{1}{L(\mu=\Lambda)}
   +\frac{2}{T(\mu=\Lambda)}
   \right]_{p\to p_{12},p_{21}},
\label{eq:(4.1)}
\end{align}
where the functions $L_{11}(\mu)$, $T_{11}(\mu)$ (contained in the photon vacuum
polarization), $L(\mu)$, and~$T(\mu)$ (contained in the W-boson vacuum
polarization) are given by~Eq.~\eqref{eq:(2.14)}; here, we have explicitly
written the dependence on the renormalization scale~$\mu$. To derive the last
expression, we have noted
\begin{equation}
   1-\frac{\beta_0\lambda}{16\pi^2}\ln\left(\frac{e^{5/3}\mu^2}{p^2}\right)
   =-\frac{\beta_0\lambda}{16\pi^2}\ln\left(\frac{e^{5/3}\Lambda^2}{p^2}\right)
\label{eq:(4.2)}
\end{equation}
for the one-loop dynamical scale~\eqref{eq:(2.9)}. (The renormalization scale
of the coupling~$\lambda$ is always set to~$\mu$ even in~$L_{11}(\mu=\Lambda)$
etc. In these expressions, we mean that only the argument of the logarithm is
set to~$\mu=\Lambda$.) We note that in the last expression
of~Eq.~\eqref{eq:(4.1)}, the overall factor~$\lambda=\lambda(\mu^2)$ is
actually canceled against the overall factor~$\lambda=\lambda(\mu^2)$ of the
functions $L_{\ell r}(\mu=\Lambda)$, $T_{\ell r}(\mu=\Lambda)$, \dots\ in the
denominator, and thus the gluon condensate~\eqref{eq:(4.1)} is clearly renormalization
group invariant. In~Fig.~\ref{fig:3}, we plot the functions appearing in the
denominators of~Eq.~\eqref{eq:(4.1)}, i.e.,
\begin{equation}
   \frac{16\pi^2}{\beta_0\lambda}
   \begin{cases}
   \left.L_{11}(\mu=\Lambda)\right|_{p_3=0},\\
   \left.T_{11}(\mu=\Lambda)\right|_{p_3=0},\\
   \left.L(\mu=\Lambda)\right|_{p_3=1/(2R)},\\
   \left.T(\mu=\Lambda)\right|_{p_3=1/(2R)},\\
   \end{cases}
\label{eq:(4.3)}
\end{equation}
as functions of~$\bm{p}^2\equiv\sum_{i=0}^2p_i^2$ for various values of the
compactification radius~$R\Lambda$. As already noted, these quantities are
renormalization group invariant. Here, $p_3$ is set to the smallest possible
values, i.e., $p_3=0$ for $L_{11}(\mu=\Lambda)$ and~$T_{11}(\mu=\Lambda)$,
and $p_3=1/(NR)=1/(2R)$ for $L(\mu=\Lambda)$ and~$T(\mu=\Lambda)$ as implied
in~Eq.~\eqref{eq:(4.1)}.

In~Fig.~\ref{fig:3b}, we see that $T_{11}(\mu=\Lambda)|_{p_3=0}$ goes to a
finite value (rather than infinity) as~$p^2=\bm{p}^2\to0$. This is precisely
the disappearance of the logarithmic factor that we observed in the previous
section; the curves in~Fig.~\ref{fig:3b} are nothing but the $N=2$ curve
in~Fig.~\ref{fig:1} up to trivial addition and rescaling. Therefore, according
to the argument in~Sect.~\ref{sec:2}, this implies the absence of a factorial
growth of perturbative coefficients and the IR renormalon ambiguity in the
Borel procedure (i.e., one considers the Borel transform of the perturbative
expansion and then performs the Borel integral).\footnote{We can also explain
the absence of the IR renormalon explicitly in the following way. For instance,
for the $T_{11}$ part, from the first expression in~Eq.~\eqref{eq:(4.1)}
and~Eq.~\eqref{eq:(3.12)}, the perturbative expansion is given by
\begin{equation}
   -\lambda\int\frac{d^3p}{(2\pi)^3}\,\frac{1}{2\pi R}\sum_{p_3}
   \left[\ln(e^{5/3}\mu^2/p^2)+f_0(p^2R^2)_{11}\right]^k
   \left(\frac{\beta_0\lambda}{16\pi^2}\right)^k
\notag
\end{equation}
and then the Borel transform~\eqref{eq:(2.5)} is obtained as
\begin{equation}
   B(u)=-\int\frac{d^3p}{(2\pi)^3}\,\frac{1}{2\pi R}\sum_{p_3}
   e^{u[\ln(e^{5/3}\mu^2/p^2)+f_0(p^2R^2)_{11}]}.
\notag
\end{equation}
Since $\ln(e^{5/3}\mu^2/p^2)+f_0(p^2R^2)_{11}\sim\text{const.}$ in the IR region,
the momentum integral (and the sum) giving the Borel transform is not IR
divergent for any~$u>0$; the Borel transform does not possess singularities
at~$u>0$. This is in contrast with the uncompactified case, where
$\ln(e^{5/3}\mu^2/p^2)+f_0(p^2R^2)_{11}$ is replaced by~$\ln(e^{5/3}\mu^2/p^2)$
(and also $\int\frac{d^3p}{(2\pi)^3}\,\frac{1}{2\pi R}\sum_{p_3}\to%
\int\frac{d^4p}{(2\pi)^4}$) and the Borel transform possesses a singularity at
certain~$u>0$.} This absence of the IR renormalon persists as far as the
compactification radius~$R\Lambda$ is finite.

However, the momentum integration of~$[T_{11}(\mu=\Lambda)]^{-1}$ (with~$p_3=0$)
itself, as given by~Eq.~\eqref{eq:(4.1)}, becomes ill defined and ambiguous
when $R\Lambda$ is finite but sufficiently large. This is caused by a simple
zero of~$T_{11}(\mu=\Lambda)$ at~$\bm{p}^2=p^2>0$, which arises when $R\Lambda$
is sufficiently large as shown in~Fig.~\ref{fig:3b}. We note that
Eq.~\eqref{eq:(4.1)} is a resummed quantity of the perturbative series in a
different way from the Borel procedure, which is obtained with resummation of a
geometric series. Similarly, $T_{11}(\mu=\Lambda)$ with other discrete values
of~$p_3$ can possess a zero as the function of~$\bm{p}^2$ and then the momentum
integration becomes ambiguous; these ambiguities are what we term the
renormalon precursor. This is not the conventional renormalon because
$T_{11}(\mu=\Lambda)|_{p_3=0}$ has no logarithmic factor~$\sim\ln p^2$
as~$p^2=\bm{p}^2\to0$ and thus the perturbative coefficients do not exhibit
factorial growth. However, instead, the momentum integration becomes ambiguous.
This is a ``precursor'' of the IR renormalon in the sense that, under the
decompactification~$R\Lambda\to\infty$, the sum of ambiguities arising from
each zero of~$T_{11}(\mu=\Lambda)$ (corresponding to different discrete values
of~$p_3$) smoothly reduces to the IR renormalon ambiguity in~$\mathbb{R}^4$.
(In an example in~Appendix~\ref{sec:B}, this summation over ambiguities is
explicitly calculated.) To see this, we note that, as~$R\Lambda\to\infty$, the
terms containing the modified Bessel function in~Eq.~\eqref{eq:(2.14)} are
suppressed for any finite~$p^2>0$ (see~Eqs.~\eqref{eq:(3.45)}
and~\eqref{eq:(3.46)}) and thus
\begin{equation}
   T_{11}(\mu=\Lambda)\stackrel{R\Lambda\to\infty}{\to}
   \frac{\beta_0\lambda}{16\pi^2}
   \ln\left(\frac{e^{5/3}\Lambda^2}{p^2}\right),
\label{eq:(4.4)}
\end{equation}
as shown in Fig.~\ref{fig:3b}, where the dashed line corresponds to the
right-hand side of this equation. We thus have
\begin{equation}
   -\lambda\int\frac{d^3p}{(2\pi)^3}\,\frac{1}{2\pi R}\sum_{p_3}
   \frac{1}{T_{11}(\mu=\Lambda)}
   \stackrel{R\Lambda\to\infty}{\to}
   -\lambda\int\frac{d^4p}{(2\pi)^4}\,
   \frac{1}{\dfrac{\beta_0\lambda}{16\pi^2}
   \ln\left(\dfrac{e^{5/3}\Lambda^2}{p^2}\right)}.
\label{eq:(4.5)}
\end{equation}
The integrand of the momentum integration in~Eq.~\eqref{eq:(4.5)} possesses the
pole singularity at~$p^2=e^{5/3}\Lambda^2$ as
\begin{equation}
   \sim\frac{16\pi^2}{\beta_0}\int\frac{d^4p}{(2\pi)^4}\,
   \frac{e^{5/3}\Lambda^2}{p^2-e^{5/3}\Lambda^2}
   =\frac{1}{\beta_0}\int_0^\infty d(p^2)\,p^2
   \frac{e^{5/3}\Lambda^2}{p^2-e^{5/3}\Lambda^2}.
\label{eq:(4.6)}
\end{equation}
The ambiguity of this momentum integration (the renormalon precursor) in the
$R\to\infty$ limit gives rise to an ambiguity:
\begin{equation}
   \pm i\pi\frac{e^{10/3}}{\beta_0}\Lambda^4.
\label{eq:(4.7)}
\end{equation}
We have defined the ambiguity by the imaginary part that appears when the
integration contour is deformed in the complex $p^2$-plane such that it avoids
the pole.

This ambiguity of the renormalon precursor in the $R\to\infty$ limit is exactly
the same as the renormalon ambiguity
in~$\mathbb{R}^4$.\footnote{In~$\mathbb{R}^4$, the IR renormalon ambiguity can
be viewed as the ambiguity arising from the momentum
integration~\cite{Novikov:1984rf} as well as the ambiguity in the Borel
integral. The renormalon precursor is thus analogous to the former picture; the
renormalon precursor however does not always coincide with the ambiguity in the
Borel integral.} In~$\mathbb{R}^4$, the corresponding part and its perturbative
expansion are given by
\begin{equation}
   -\lambda\int\frac{d^4p}{(2\pi)^4}\,
   \frac{1}{\dfrac{\beta_0\lambda}{16\pi^2}
   \ln\left(\dfrac{e^{5/3}\Lambda^2}{p^2}\right)}
   \stackrel{\text{expansion in $\lambda$}}{\to}
   \lambda\sum_{k=0}^\infty
   f_k\left(\frac{\beta_0\lambda}{16\pi^2}\right)^k
\label{eq:(4.8)}
\end{equation}
with
\begin{equation}
   f_k=\int\frac{d^4p}{(2\pi)^4}\,
   \left[\ln\left(\frac{e^{5/3}\mu^2}{p^2}\right)\right]^k
   \stackrel{k\to\infty}{\sim}\frac{e^{10/3}\mu^4}{16\pi^2}\frac{k!}{2^{k+1}},
\label{eq:(4.9)}
\end{equation}
where we have used Eq.~\eqref{eq:(4.2)} in the perturbative expansion. This
factorial growth of the perturbative coefficients produces the pole in the
Borel transform~\eqref{eq:(2.5)} $-e^{10/3}\mu^4/(16\pi^2)1/(u-2)$ and, through
the Borel integral~\eqref{eq:(2.7)}, the IR renormalon ambiguity
\begin{equation}
   \pm i\pi\frac{e^{10/3}}{\beta_0}\Lambda^4.
\label{eq:(4.10)}
\end{equation}
This is the same as~Eq.~\eqref{eq:(4.7)}.

The situation is similar for~$L(\mu=\Lambda)$ and~$T(\mu=\Lambda)$
in~Eq.~\eqref{eq:(4.1)}. As far as~$R\Lambda$ is finite,
$L(\mu=\Lambda)$ and~$T(\mu=\Lambda)$ in~Eq.~\eqref{eq:(4.1)} do not diverge
as~$\bm{p}^2\to0$, because of the twisted momentum, $p_3=n/R+1/(2R)\neq0$
for~$n\in\mathbb{Z}$. Thus, there is neither logarithmic factor nor IR
renormalon. On the other hand, it can be shown that the terms containing the
modified Bessel function in~Eq.~\eqref{eq:(2.14)} are suppressed for any
finite~$p$ as~$R\Lambda\to\infty$~\cite{Ashie:2019cmy} and
\begin{equation}
   L(\mu=\Lambda),T(\mu=\Lambda)
   \stackrel{R\Lambda\to\infty}{\to}
   \frac{\beta_0\lambda}{16\pi^2}
   \ln\left(\frac{e^{5/3}\Lambda^2}{p^2}\right).
\label{eq:(4.11)}
\end{equation}
According to these behaviors, $L(\mu=\Lambda)$ and~$T(\mu=\Lambda)$ acquire
zeros as~$R \Lambda$ becomes larger (but still finite) and the momentum
integrations of~$L(\mu=\Lambda)^{-1}$ and~$T(\mu=\Lambda)^{-1}$ become
ambiguous; the presence of these zeros can be seen in~Figs.~\ref{fig:3c}
and~\ref{fig:3d}. The ambiguity caused by these zeros (i.e., the renormalon
precursor) coincides with the IR renormalon ambiguity as~$R\Lambda\to\infty$ as
in the above case of~$T_{11}(\mu=\Lambda)$.

The situation is slightly different for~$L_{11}(\mu=\Lambda)$. As shown
in~Eqs.~\eqref{eq:(3.32)} and~\eqref{eq:(3.35)}, $L_{11}(\mu=\Lambda)|_{p_3=0}$
has a more singular behavior for~$p^2=\bm{p}^2\to0$ as
\begin{align}
   \left.L_{11}(\mu=\Lambda)\right|_{p_3=0}
   &=\frac{\beta_0\lambda}{16\pi^2}
   \left[
   \frac{1}{p^2R^2}
   +\ln(\Lambda^2R^2)+2\gamma+\ln4-1
   +O(p^2R^2\ln(p^2R^2))
   \right]
\notag\\
   &=
   -\frac{1}{p^2}\left[
   (1+O(\lambda))p^2+m_{\text{sc}}^2
   +O(p^4R^2\ln(p^2R^2))
   \right]
\label{eq:(4.12)}
\end{align}
and $L_{11}(\mu=\Lambda)|_{p_3=0}$ diverges as~$p^2=\bm{p}^2\to0$ as clearly seen
in~Fig.~\ref{fig:3a}. Here, the screening mass~$m_{\text{sc}}$ has been
introduced by~$m_{\text{sc}}^2\equiv-\beta_0\lambda/(16\pi^2R^2)$. As noted in
footnote~\ref{footnote:6}, in perturbative expansion, we do not regard
$m_{\text{sc}}$ as an $\mathcal{O}(\lambda)$ quantity and treat it as if it was
an $\mathcal{O}(\lambda^0)$ quantity, to avoid IR divergences in fixed-order
perturbation theory. In this treatment, the situation becomes similar to the
above cases, and it is concluded that there is no IR renormalon. On the other
hand, again in~$L_{11}(\mu=\Lambda)$ the terms containing the modified Bessel
function in~Eq.~\eqref{eq:(2.14)} are suppressed for fixed $p^2>0$
as~$R\Lambda\to\infty$:
\begin{equation}
   L_{11}(\mu=\Lambda)
   \stackrel{R\Lambda\to\infty}{\to}
   \frac{\beta_0\lambda}{16\pi^2}
   \ln\left(\frac{e^{5/3}\Lambda^2}{p^2}\right).
\label{eq:(4.13)}
\end{equation}
Thus, the ambiguity in the momentum integration
of~$L_{11}(\mu=\Lambda)|_{p_3=0}^{-1}$, i.e., the renormalon precursor, caused by
the zero in~Fig.~\ref{fig:3a} is smoothly reduced to the IR renormalon
ambiguity in~$\mathbb{R}^4$ as $R\Lambda\to\infty$.

Although in this section we have only demonstrated the presence of the
renormalon precursor in the $N=2$ theory, this notion must be quite general,
being applicable to QCD(adj.) with any~$N$. See also Appendix~\ref{sec:B} for
other examples where this notion applies.

\section{Conclusion}
\label{sec:5}
In this paper, we made some remarks on the issue of the possible existence of
the IR renormalon in the $SU(N)$ QCD(adj.) on~$\mathbb{R}^3\times S^1$ with the
$\mathbb{Z}_N$ twisted boundary condition, by making use of the large-$\beta_0$
approximation. In the first part of this paper, we showed that for any
finite~$N$ the photon vacuum polarization loses the logarithmic
factor~$\ln p^2$ as~$p^2\to0$ and there is no IR renormalon in the compactified
spacetime~$\mathbb{R}^3\times S^1$. In the second part, we presented the notion
of the renormalon precursor, i.e., the ambiguity in the momentum integration,
that smoothly reduces to the IR renormalon ambiguity in~$\mathbb{R}^4$ under
the decompactification~$\mathbb{R}^3\times S^1\to\mathbb{R}^4$. On the first
issue, although our demonstration of the disappearance of the logarithmic
factor required very detailed calculations, there might be a more direct and
simpler way to understand the absence of the logarithmic factor. On the second
issue, the renormalon precursor is a quite general notion as an object which
smoothly complements the difference between the absence and existence of the IR
renormalon under the removal of an IR cutoff (such as the compactification
radius, the mass, etc.).

\section*{Acknowledgements}
This work was supported by JSPS Grant-in-Aid for Scientific Research Grant
Numbers JP18J20935 (O.M.), JP16H03982, JP20H01903 (H.S.),
and~JP19K14711 (H.T.).

\appendix

\section{Rigorous proof of Eqs.~\eqref{eq:(3.29)}--\eqref{eq:(3.31)} up
to~$O((p^2R^2)^0)$}
\label{sec:A}
In this appendix, we give a rigorous proof for the asymptotic expansion
in~Eqs.~\eqref{eq:(3.29)}--\eqref{eq:(3.31)} up to~$O((p^2R^2)^0)$ terms. This
is sufficient to conclude the disappearance of the logarithmic factor~$\ln p^2$
as~$p^2R^2\to0$ in the photon vacuum polarization and the absence of the IR
renormalon.

We study the function~\eqref{eq:(3.11)},
\begin{equation}
   f_\nu(p^2R^2)_{\ell r}
   =24\sum_{j=1}^\infty(\sigma_{j,N})_{\ell r}
   \int_0^1dx\,x(1-x)K_\nu(2j\Hat{p}(x))
\label{eq:(A1)}
\end{equation}
for $\nu=0$ and~$2$, where we have
set~$\Hat{p}(x)\equiv\pi\sqrt{x(1-x)}\sqrt{p^2R^2}$. First, we note that
$(\sigma_{j,N})_{\ell r}$ in~Eq.~\eqref{eq:(2.16)} can be represented as
\begin{equation}
   (\sigma_{j,N})_{\ell r}=\sum_{m,n=1}^NC_{\ell r}^{mn}e^{i(n-m)2\pi j/N}
\label{eq:(A2)}
\end{equation}
with 
\begin{equation}
   C_{\ell r}^{mn}\equiv\frac{1}{N}(\nu^m-\nu^n)_\ell(\nu^m-\nu^n)_r.
\label{eq:(A3)}
\end{equation}
Then, using integral representations of the modified Bessel functions,
\begin{equation}
   K_0(z)=\int_0^1\frac{dt}{t}\,e^{-\frac{z}{2}(t+\frac{1}{t})},\qquad
   K_2(z)=\frac{1}{2}\int_0^1\frac{dt}{t}\,
   \left(t^2+\frac{1}{t^2}\right)e^{-\frac{z}{2}(t+\frac{1}{t})},
\label{eq:(A4)}
\end{equation}
we obtain
\begin{align}
   f_0(p^2R^2)_{\ell r}
   &=24\sum_{m,n=1}^NC_{\ell r}^{mn}
   \lim_{j_{\text{max}}\to\infty}\sum_{j=1}^{j_{\text{max}}}
   e^{i(n-m)2\pi j/N}\int_0^1dx\,x(1-x)
   \int_0^1\frac{dt}{t}\,e^{-j\Hat{p}(x)(t+\frac{1}{t})}
\notag\\
   &=24\sum_{m,n=1}^NC_{\ell r}^{mn}
   \lim_{j_{\text{max}}\to\infty}
   \int_0^1dx\,x(1-x) 
   \int_0^1\frac{dt}{t}\,\sum_{j=1}^{j_{\text{max}}}
   \left[e^{i(n-m)2\pi/N}e^{-\Hat{p}(x)(t+\frac{1}{t})}\right]^j
\notag\\
   &=24\sum_{m,n=1}^NC_{\ell r}^{mn}e^{i(n-m)2\pi/N}
   \int_0^1dx\,x(1-x)
   \int_0^1\frac{dt}{t}\,
   \frac{1}{e^{\Hat{p}(x)(t+\frac{1}{t})}-e^{i(n-m)2\pi/N}}
\notag\\
   &\qquad{}
   -24\sum_{m,n=1}^NC_{\ell r}^{mn}e^{i(n-m)2\pi/N}
\notag\\
   &\qquad\qquad\qquad{}
   \times\lim_{j_{\text{max}}\to\infty}
   \int_0^1dx\,x(1-x)
   \int_0^1\frac{dt}{t}\,
   \frac{[e^{i(n-m)2\pi/N}e^{-\Hat{p}(x)(t+\frac{1}{t})}]^{j_{\text{max}}}}
   {e^{\Hat{p}(x)(t+\frac{1}{t})}-e^{i(n-m)2\pi/N}}.
\label{eq:(A5)}
\end{align}
The second term in the last line vanishes because
\begin{align}
   &\left|\int_0^1dx\,x(1-x)
   \int_0^1\frac{dt}{t}\,
   \frac{[e^{i(n-m)2\pi/N}e^{-\Hat{p}(x)(t+\frac{1}{t})}]^{j_{\text{max}}}}
   {e^{\Hat{p}(x)(t+\frac{1}{t})}-e^{i(n-m)2\pi/N}}\right|
\notag\\
   &\leq\int_0^1dx\,x(1-x)
   \int_0^1\frac{dt}{t}\,
   \frac{e^{-j_{\text{max}}\Hat{p}(x)(t+\frac{1}{t})}}
   {|e^{\Hat{p}(x)(t+\frac{1}{t})}-e^{i(n-m)2\pi/N}|}
\notag\\
   &<\frac{1}{\left|\Im e^{i(n-m)2\pi/N}\right|}
   \int_0^1dx\,x(1-x)\int_0^1\frac{dt}{t}\,
   e^{-j_{\text{max}}\Hat{p}(x)(t+\frac{1}{t})}
\notag\\
   &<\frac{1}{j_{\text{max}}}
   \frac{1}{\left|\Im e^{i(n-m)2\pi/N}\right|}
   \int_0^1dx\,x(1-x)\frac{1}{\Hat{p}(x)},
\label{eq:(A6)}
\end{align}
where, in the last step, we have used
\begin{equation}
   \int_0^1\frac{dt}{t}\,e^{-j_{\text{max}}\Hat{p}(x)(t+\frac{1}{t})}
   =K_0(2j_{\text{max}}\Hat{p}(x))
   <\frac{1}{j_{\text{max}}\Hat{p}(x)}e^{-j_{\text{max}}\Hat{p}(x)}
   <\frac{1}{j_{\text{max}}\Hat{p}(x)}
\label{eq:(A7)}
\end{equation}
for~$\Hat{p}(x)>0$, as shown in~Eq.~(B3) of~Ref.~\cite{Ishikawa:2019tnw}.
Hence, we obtain
\begin{equation}
   f_0(p^2R^2)_{\ell r}
   =24\sum_{m,n=1}^NC_{\ell r}^{mn}e^{i(n-m)2\pi/N}
   \int_0^1dx\,x(1-x)
   \int_0^1\frac{dt}{t}\,
   \frac{1}{e^{\Hat{p}(x)(t+\frac{1}{t})}-e^{i(n-m)2\pi/N}},
\label{eq:(A8)}
\end{equation}
and, in a similar manner,\footnote{For~$f_2(p^2R^2)_{\ell r}$, one can use
$K_2(2z)<[(1/z)+(1/z)^2]e^{-z}$, which is shown in~Eq.~(B5)
of~Ref.~\cite{Ashie:2019cmy}.}
\begin{align}
   &f_2(p^2R^2)_{\ell r}
\notag\\
   &=12\sum_{m,n=1}^NC_{\ell r}^{mn}e^{i(n-m)2\pi/N}
   \int_0^1dx\,x(1-x)
   \int_0^1\frac{dt}{t}\,
   \left(t^2+\frac{1}{t^2}\right)
   \frac{1}{e^{\Hat{p}(x)(t+\frac{1}{t})}-e^{i(n-m)2\pi/N}}.
\label{eq:(A9)}
\end{align}

To obtain the asymptotic behavior of~$f_0(p^2R^2)_{\ell r}$
from~Eq.~\eqref{eq:(A8)}, we show
\begin{align}
   &\int_0^1\frac{dt}{t}\,
   \frac{1}{e^{\Hat{p}(x)(t+\frac{1}{t})}-c}
   -\left[-\frac{1}{1-c}\ln\Hat{p}(x)\right]
\notag\\ 
   &=\int_0^{\Hat{p}(x)}\frac{dt}{t}\,
   \frac{1}{e^{\Hat{p}(x)(t+\frac{1}{t})}-c}
   +\int_{\Hat{p}(x)}^1\frac{dt}{t}\,
   \left[\frac{1}{e^{\Hat{p}(x)(t+\frac{1}{t})}-c}-\frac{1}{1-c}\right]
\notag\\
   &=O(\Hat{p}(x)^0)
\label{eq:(A10)}
\end{align}
for small $\Hat{p}(x)>0$;\footnote{We note that
\begin{align}
   &\int_0^1dx\,x(1-x)\int_0^1\frac{dt}{t}\,
   \frac{1}{e^{\Hat{p}(x)(t+\frac{1}{t})}-e^{i(n-m)2\pi/N}}
\notag\\
   &=\lim_{\delta_1,\delta_2\to 0}\int_{\delta_1}^{1-\delta_2}dx\,x(1-x)
   \int_0^1\frac{dt}{t}\,
   \frac{1}{e^{\Hat{p}(x)(t+\frac{1}{t})}-e^{i(n-m)2\pi/N}},
\notag
\end{align}
and $\Hat{p}(x)$ can be assumed to be positive.} here and hereafter, we set
$c\equiv e^{i(n-m)2\pi/N}\neq1$. Then, Eq.~\eqref{eq:(A10)} tells us that
\begin{equation}
   \int_0^1\frac{dt}{t}\,\frac{1}{e^{\Hat{p}(x)(t+\frac{1}{t})}-c}
   =-\frac{1}{1-c}\ln\Hat{p}(x)+O(\Hat{p}(x)^0).
\label{eq:(A11)}
\end{equation}
Now, for the first term in the second line of~Eq.~\eqref{eq:(A10)}, we have
\begin{align}
   \left|\int_0^{\Hat{p}(x)}\frac{dt}{t}\,
   \frac{1}{e^{\Hat{p}(x)(t+\frac{1}{t})}-c}\right|
   &\leq\int_0^{\Hat{p}(x)}\frac{dt}{t}\,
   \frac{1}{|e^{\Hat{p}(x)(t+\frac{1}{t})}-c|}
\notag\\
   &<\int_0^{\Hat{p}(x)}\frac{dt}{t}\,
   \frac{1}{e^{\Hat{p}(x)(t+\frac{1}{t})}-1}
\notag\\
   &<\int_0^1\frac{dt}{t}\,\frac{1}{e^{1/t}-1}
   =O(\Hat{p}(x)^0).
\label{eq:(A12)}
\end{align}
Here, we have used $e^{\Hat{p}(x)(t+1/t)}-1\geq e^{\Hat{p}(x)/t}-1$. The second
term in the second line of~Eq.~\eqref{eq:(A10)} can be bounded as
\begin{equation}
   \left|\int_{\Hat{p}(x)}^1\frac{dt}{t}\,
   \frac{e^{\Hat{p}(x)(t+\frac{1}{t})}-1}
   {[e^{\Hat{p}(x)(t+\frac{1}{t})}-c](1-c)}\right|
   \leq\frac{1}{\left|\Im c\right|^2}
   \int_{\Hat{p}(x)}^1\frac{dt}{t}\,
   \left[e^{\Hat{p}(x)(t+\frac{1}{t})}-1\right].
\label{eq:(A13)}
\end{equation}
Using, for instance,\footnote{We assume that $p^2R^2$ is small enough such that
$\Hat{p}(x)\leq1$ is satisfied for~$0\leq\forall x\leq1$.}
\begin{equation}
   e^{\Hat{p}(x)(t+\frac{1}{t})}-1
   \leq8\Hat{p}(x)\left(t+\frac{1}{t}\right),\qquad
   \text{for $\Hat{p}(x)\leq t\leq1$},
\label{eq:(A14)}
\end{equation}
we can show
\begin{equation}
   \left|\int_{\Hat{p}(x)}^1\frac{dt}{t}\,
   \frac{e^{\Hat{p}(x)(t+\frac{1}{t})}-1}
   {[e^{\Hat{p}(x) (t+\frac{1}{t})}-c](1-c)}\right| 
   =O(\Hat{p}(x)^0).
\label{eq:(A15)}
\end{equation}
Equations \eqref{eq:(A12)} and~\eqref{eq:(A15)} show Eq.~\eqref{eq:(A10)} and
thus Eq.~\eqref{eq:(A11)}.

We now study $f_2(p^2R^2)_{\ell r}$ in~Eq.~\eqref{eq:(A9)}. We first note
\begin{align}
   \int_0^1\frac{dt}{t}\,
   \left(t^2+\frac{1}{t^2}\right)\frac{1}{e^{\Hat{p}(x)(t+\frac{1}{t})}-c} 
   =\int_1^\infty\frac{ds}{\sqrt{s^2-1}}\,
   (4s^2-2)\frac{1}{e^{2\Hat{p}(x)s}-c},
\label{eq:(A16)}
\end{align}
under the change of variable, $2s=t+1/t$. In the following we prove, for the
right-hand side,
\begin{align}
   &\int_1^\infty\frac{ds}{\sqrt{s^2-1}}\,
   (4s^2-2)\frac{1}{e^{2\Hat{p}(x)s}-c}
   -\int_0^\infty ds\,\frac{4s}{e^{2\Hat{p}(x)s}-c}
\notag\\
   &=\int_1^\infty ds\,
   \left(\frac{4s^2-2}{\sqrt{s^2-1}}-4s\right)\frac{1}{e^{2\Hat{p}(x)s}-c}
   -\int_0^1ds\,\frac{4s}{e^{2\Hat{p}(x)s}-c}
\notag\\
   &=O(\Hat{p}(x)^0).
\label{eq:(A17)}
\end{align}
Then, we obtain
\begin{equation}
   \int_1^\infty\frac{ds}{\sqrt{s^2-1}}\,
   (4s^2-2)\frac{1}{e^{2\Hat{p}(x)s}-c}
   =\frac{\Li_2(c)}{c}
   \frac{1}{\Hat{p}(x)^2}+O(\Hat{p}(x)^0),
\label{eq:(A18)}
\end{equation}
by noting that the polylogarithm function~\eqref{eq:(3.24)} can be represented
as
\begin{equation}
   \Li_2(c)=c\Hat{p}(x)^2\int_0^{\infty}ds\,\frac{4s}{e^{2\Hat{p}(x)s}-c}.
\label{eq:(A19)}
\end{equation}

Now we show Eq.~\eqref{eq:(A17)}. For the first term in Eq.~\eqref{eq:(A17)},
using
\begin{equation}
   \frac{4s^2-2}{\sqrt{s^2-1}}-4s
   \leq\frac{4}{\sqrt{s-1}}\frac{1}{s^2},
\label{eq:(A20)}
\end{equation}
we obtain
\begin{align}
   \left|\int_1^\infty ds\,
   \left(\frac{4s^2-2}{\sqrt{s^2-1}}-4s\right)
   \frac{1}{e^{2\Hat{p}(x)s}-c}\right|
   &\leq\int_1^\infty ds\,
   \frac{4}{\sqrt{s-1}}\frac{1}{s^2}\left|
   \frac{1}{e^{2\Hat{p}(x)s}-c}\right|
\notag\\
   &<\frac{1}{\left|\Im c\right|}\int_1^\infty ds\,
  \frac{4}{\sqrt{s-1}}\frac{1}{s^2}
   =O(\Hat{p}(x)^0).
\label{eq:(A21)}
\end{align}
For the second term in~Eq.~\eqref{eq:(A17)}, we immediately obtain
\begin{align}
   \left|\int_0^1ds\,
   \frac{4s}{e^{2\Hat{p}(x)s}-c}\right|
   \leq\int_0^1ds\,
   \left|\frac{4s}{e^{2\Hat{p}(x)s}-c}\right|
   \leq\frac{1}{\left|\Im c\right|}\int_0^1ds\,4s=O(\Hat{p}(x)^0).
\label{eq:(A22)}
\end{align}
Hence, Eq.~\eqref{eq:(A17)} and thus Eq.~\eqref{eq:(A18)} have been shown.

Finally, from Eqs.~\eqref{eq:(A8)} and~\eqref{eq:(A11)}, we obtain
\begin{align}
   f_0(p^2R^2)_{\ell r}
   &=-2\sum_{m,n=1}^NC_{\ell r}^{mn}
   \frac{e^{i(n-m)2\pi/N}}{1-e^{i(n-m)2\pi/N}}
   \ln(p^2R^2)+O((p^2R^2)^0)
\notag\\
   &=\frac{2}{N}\sum_{j=1}^Nj(\sigma_{j,N})_{\ell r}
   \ln(p^2R^2)+O((p^2R^2)^0)
\notag\\
   &=\delta_{\ell r}\ln(p^2R^2)+O((p^2R^2)^0),
\label{eq:(A23)}
\end{align}
where we have used $e^{i(n-m)2\pi/N}/[1-e^{i(n-m)2\pi/N}]=%
-\frac{1}{N}\sum_{j=1}^Nje^{i(n-m)2\pi j/N}$ and~Eq.~\eqref{eq:(3.8)}.

Also, from~Eqs.~\eqref{eq:(A9)}, \eqref{eq:(A16)}, and~\eqref{eq:(A18)}, we
obtain
\begin{align}
   f_2(p^2R^2)_{\ell r}
   &=\frac{12}{\pi^2}\sum_{m,n=1}^NC_{\ell r}^{mn}
   \Li_2(e^{i(n-m)2\pi/N})\frac{1}{p^2R^2}+O((p^2R^2)^0)
\notag\\
   &=\frac{12}{\pi^2}\sum_{j=1}^\infty(\sigma_{j,N})_{\ell r}
   \frac{1}{j^2}\frac{1}{p^2 R^2}+O((p^2R^2)^0).
\label{eq:(A24)}
\end{align}
These results (Eqs.~\eqref{eq:(A23)} and~\eqref{eq:(A24)}) prove the asymptotic
expansion in~Eqs.~\eqref{eq:(3.29)}--\eqref{eq:(3.31)} up to~$O((p^2R^2)^0)$.

\section{Renormalon precursor in wider context}
\label{sec:B}
In this appendix, we present some other examples to which the notion of the
renormalon precursor applies.

Our first example is the understanding of the shift of the Borel singularity
by~$-1/2$ under the
compactification~$\mathbb{R}^d\to\mathbb{R}^{d-1}\times S^1$~\cite{Ishikawa:2019oga} and its relation to the decompactification limit. We start with the
integral in~$\mathbb{R}^d$:
\begin{equation}
   \mathcal{I}(\alpha;d)
   \equiv\int\frac{d^dp}{(2\pi)^d}\,(p^2)^\alpha\lambda(p^2),
\label{eq:(B1)}
\end{equation}
which provides a typical example where we have an IR renormalon; here,
$\lambda(p^2)$ is the one-loop running coupling:
\begin{equation}
   \lambda(p^2)\equiv\frac{(4\pi)^{d/2}}{\beta_0}\frac{1}{\ln(p^2/\Lambda^2)},
   \qquad
   \Lambda^2\equiv\mu^2e^{-(4\pi)^{d/2}/[\beta_0\lambda(\mu^2)]}.
\label{eq:(B2)}
\end{equation}
Then by noting
\begin{equation}
   \lambda(p^2)=\lambda(\mu^2)\sum_{k=0}^\infty
   \left[\ln\left(\frac{\mu^2}{p^2}\right)\right]^k
   \left[\frac{\beta_0\lambda(\mu^2)}{(4\pi)^{d/2}}\right]^k,
\label{eq:(B3)}
\end{equation}
the perturbative expansion of~Eq.~\eqref{eq:(B1)} is given by
\begin{equation}
   \mathcal{I}(\alpha;d)
   \sim\lambda(\mu^2)\sum_{k=0}^\infty f_k
   \left[\frac{\beta_0\lambda(\mu^2)}{(4\pi)^{d/2}}\right]^k,\qquad
   f_k=\int\frac{d^dp}{(2\pi)^d}\,(p^2)^\alpha
   \left[\ln\left(\frac{\mu^2}{p^2}\right)\right]^k.
\label{eq:(B4)}
\end{equation}
The corresponding Borel transform (see~Eq.~\eqref{eq:(2.5)}) is
\begin{align}
   B[\mathcal{I}(\alpha,d)](u)
   &=\int\frac{d^dp}{(2\pi)^d}\,(p^2)^\alpha
   \left(\frac{\mu^2}{p^2}\right)^u
\notag\\
   &=\mu^{2u}\frac{1}{(4\pi)^{d/2}}\frac{1}{{\mit\Gamma}(d/2)}
   \frac{q^{2\alpha+d-2u}}{\alpha+d/2-u},
\label{eq:(B5)}
\end{align}
where we have introduced an ultraviolet cutoff~$q$, i.e., $p^2<q^2$. This
Borel transform possesses a simple pole at~$u=\alpha+d/2$ and thus the Borel
integral
\begin{equation}
   \frac{(4\pi)^{d/2}}{\beta_0}\int_0^\infty du\,
   B[\mathcal{I}(\alpha,d)](u)\,e^{-(4\pi)^{d/2}u/[\beta_0\lambda(\mu^2)]}
\label{eq:(B6)}
\end{equation}
has the ambiguity (the IR renormalon ambiguity)
\begin{equation}
   \pm i\pi\frac{1}{\beta_0}\frac{1}{{\mit\Gamma}(d/2)}\Lambda^{2\alpha+d}.
\label{eq:(B7)}
\end{equation}

Now, let us consider the $S^1$~compactification,
$\mathbb{R}^d\to\mathbb{R}^{d-1}\times S^1$ and suppose that the integrand does
not change under this compactification. Moreover, let us suppose that the KK
momentum~$p_{d-1}$ is simply given by~$p_{d-1}=n/R$ with~$n\in\mathbb{Z}$
(rather than the twisted momentum). This situation can occur for instance in
the large-$N$ limit of the 2D $\mathbb{C}P^{N-1}$ models defined on the
compactified
spacetime~$\mathbb{R}^2\to\mathbb{R}\times S^1$~\cite{Ishikawa:2019tnw,%
Ishikawa:2019oga}. Under this situation, the integral~\eqref{eq:(B1)} is
replaced by
\begin{align}
   \mathcal{I}^C(\alpha;d)
   &\equiv\int\frac{d^{d-1}p}{(2\pi)^{d-1}}\,\frac{1}{2\pi R}\sum_{p_{d-1}}
   (p^2)^\alpha\lambda(p^2)
\notag\\
   &=\int\frac{d^{d-1}p}{(2\pi)^{d-1}}\,\frac{1}{2\pi R}\sum_{p_{d-1}}
   (p^2)^\alpha\frac{(4\pi)^{d/2}}{\beta_0}\frac{1}{\ln(p^2/\Lambda^2)}.
\label{eq:(B8)}
\end{align}
As noted in~Sect.~\ref{sec:2}, for the factorial growth of perturbative
coefficients, the presence of the logarithmic behavior of the integrand
as~$p^2=\bm{p}^2+p_{d-1}^2\equiv\sum_{i=0}^{d-2}p_i^2+p_{d-1}^2\to0$ is crucial.
Then, it is sufficient to focus on the contribution with~$p_{d-1}=0$,
\begin{equation}
   \int\frac{d^{d-1}p}{(2\pi)^{d-1}}\,\frac{1}{2\pi R}
   \left.(p^2)^\alpha\lambda(p^2)\right|_{p_{d-1}=0}
\label{eq:(B9)}
\end{equation}
to detect the IR renormalon. Since this is $\mathcal{I}(\alpha,d-1)/(2\pi R)$,
the Borel transform associated with this $p_{d-1}=0$ contribution is given,
from~Eq.~\eqref{eq:(B5)}, by
\begin{equation}
   \frac{\mu^{2u}}{2\pi R}\frac{1}{(4\pi)^{(d-1)/2}}
   \frac{1}{{\mit\Gamma}((d-1)/2)}
   \frac{q^{2\alpha+d-1-2u}}{\alpha+(d-1)/2-u}.
\label{eq:(B10)}
\end{equation}
Thus, the location of the Borel singularity is shifted by~$-1/2$ under the
compactification~\cite{Ishikawa:2019oga} and the associated IR renormalon
ambiguity in~Eq.~\eqref{eq:(B6)} is, instead of~Eq.~\eqref{eq:(B7)},
\begin{equation}
   \pm i\pi\frac{1}{\beta_0}\frac{\sqrt{4\pi}}{{\mit\Gamma}((d-1)/2)}
   \frac{\Lambda^{2\alpha+d-1}}{2\pi R}.
\label{eq:(B11)}
\end{equation}
This is the IR renormalon ambiguity
in~$\mathcal{I}^C(\alpha;d)$~\eqref{eq:(B8)}. Thus, under the
decompactification~$R\to\infty$, the behavior of the IR renormalon ambiguity
suddenly changes; Eq.~\eqref{eq:(B11)} vanishes as~$R\to\infty$
and~Eq.~\eqref{eq:(B7)} emerges suddenly at~$R=\infty$. The renormalon
precursor fills this gap and provides a smooth change under the
decompactification.

To see this, we note that when the KK momentum in~Eq.~\eqref{eq:(B8)} satisfies
\begin{equation}
   |p_{d-1}|<\Lambda
\label{eq:(B12)}
\end{equation}
the integrand of the $d-1$-dimensional momentum integral possesses a simple
pole at
\begin{equation}
   \bm{p}^2=\Lambda^2-p_{d-1}^2>0.
\label{eq:(B13)}
\end{equation}
The sum of the contributions of these poles reads (by
noting~$1/\ln(p^2/\Lambda^2)\sim\Lambda^2/(p^2-\Lambda^2)$ around the pole)
\begin{align}
   &\frac{1}{2\pi R}
   \frac{(4\pi)^{d/2}}{\beta_0}
   \int\frac{d^{d-1}p}{(2\pi)^{d-1}}\,\sum_{|p_{d-1}|<\Lambda}
   (\bm{p}^2+p_{d-1}^2)^\alpha
   \frac{\Lambda^2}{\bm{p}^2+p_{d-1}^2-\Lambda^2}
\notag\\
   &=\frac{1}{\beta_0}\frac{\sqrt{4\pi}}{{\mit\Gamma}((d-1)/2)}
   \frac{1}{2\pi R}\int_0^\infty d(\bm{p}^2)\,(\bm{p}^2)^{(d-3)/2}
   \sum_{|p_{d-1}|<\Lambda}
   (\bm{p}^2+p_{d-1}^2)^\alpha
   \frac{\Lambda^2}{\bm{p}^2+p_{d-1}^2-\Lambda^2}.
\label{eq:(B14)}
\end{align}
The sum of ambiguities arising from the momentum integrals is thus
\begin{equation}
   \pm i\pi\frac{1}{\beta_0}\frac{\sqrt{4\pi}}{{\mit\Gamma}((d-1)/2)}
   \frac{\Lambda^{2\alpha+d}}{2\pi}\frac{1}{R\Lambda}\sum_{|n|<R\Lambda}
   \left[1-\frac{n^2}{(R\Lambda)^2}\right]^{(d-3)/2}.
\label{eq:(B15)}
\end{equation}
In this sum, the $n=0$ term is the IR renormalon ambiguity in the compactified
theory, Eq.~\eqref{eq:(B11)}. Other terms in the sum are not the renormalon in
the compactified theory; their sum is what we call the renormalon precursor,
the ambiguity of the momentum integral, which does not correspond to the IR
renormalon. The total ambiguity in the compactified theory is given
by~Eq.~\eqref{eq:(B15)} and in the decompactified limit~$R\Lambda\to\infty$,
it becomes
\begin{equation}
   \stackrel{R\Lambda\to\infty}{\to}
   \pm i\pi\frac{1}{\beta_0}\frac{\sqrt{4\pi}}{{\mit\Gamma}((d-1)/2)}
   \frac{\Lambda^{2\alpha+d}}{2\pi}\int_{-1}^1 dx\,(1-x^2)^{(d-3)/2}
   =\pm i\pi\frac{1}{\beta_0}\frac{1}{{\mit\Gamma}(d/2)}
   \Lambda^{2\alpha+d},
\label{eq:(B16)}
\end{equation}
which precisely coincides with the IR renormalon ambiguity in~$\mathbb{R}^d$,
Eq.~\eqref{eq:(B7)}. In this way, by introducing the renormalon precursor, we
have a smooth transition of the ambiguity under the decompactification.

Our next example is the $SU(N)$ gauge theory in~$\mathbb{R}^4$ with massive
fermions with a degenerate mass~$m$; the mass of the fermions acts as an IR
cutoff and is analogous to the inverse of the compactification radius, $1/R$,
in the above examples. In the large-$\beta_0$ approximation,\footnote{Here, we
naively carry out the replacement~\eqref{eq:(2.1)} and do not properly take
into account the fact that the gluons are massless. In this regard, it is more
appropriate to regard the present analysis as being done in Abelian gauge
theories (with suitable modifications).} the propagator of the gauge field
reads
\begin{align}
   &\left\langle A_\mu^a(x)A_\nu^b(y)\right\rangle
\notag\\
   &=\frac{\lambda(\mu^2)}{N}\delta^{ab}\int\frac{d^4p}{(2\pi)^4}
   e^{ip(x-y)}
   \frac{1}{(p^2)^2}
   \left\{
   \left[1
   -\Pi(p^2,m^2)
   \right]^{-1}
   (p^2\delta_{\mu\nu}-p_\mu p_\nu)
   +\frac{1}{\xi}p_\mu p_\nu
   \right\},
\label{eq:(B17)}
\end{align}
where
\begin{equation}
   \Pi(p^2,m^2)=-\frac{3\beta_0\lambda(\mu^2)}{8\pi^2}
   \int_0^1 dx\,x(1-x)\ln\left[\frac{x(1-x)p^2+m^2}{\mu^2}\right].
\label{eq:(B18)}
\end{equation}
We then consider a gauge-invariant quantity~$\mathcal{F}$, which is given by
\begin{equation}
   \mathcal{F}=\lambda(\mu^2)\int\frac{d^4p}{(2\pi)^4}\,(p^2)^\alpha
   \frac{1}{1-\Pi(p^2,m^2)};
\label{eq:(B19)}
\end{equation}
see~Eq.~\eqref{eq:(2.3)}. Now, for finite $m^2$,
$\Pi(p^2,m^2)\to\text{const.}$ as~$p^2\to0$ and this does not possess a
logarithmic factor. Therefore, the perturbative expansion
of~Eq.~\eqref{eq:(B19)} does not produce a factorially divergent series or the
IR renormalon ambiguity.

Nevertheless, the momentum integral in~Eq.~\eqref{eq:(B19)} can be ill defined
and ambiguous when $m^2$ is sufficiently small. To see this, we first note
\begin{equation}
   \frac{\partial}{\partial p^2}\left[1-\Pi(p^2,m^2)\right]
   =\frac{3\beta_0\lambda(\mu^2)}{8\pi^2}
   \int_0^1dx\,\frac{\left[x(1-x)\right]^2}{x(1-x)p^2+m^2}>0,
\label{eq:(B20)}
\end{equation}
where we have assumed the asymptotic freedom~$\beta_0>0$. Therefore, the
function~$1-\Pi(p^2,m^2)$ is a monotonically increasing function of~$p^2$. On
the other hand, at the end points we have
\begin{align}
   1-\Pi(p^2=0,m^2)
   &=1+\frac{\beta_0\lambda(\mu^2)}{16\pi^2}\ln\left(\frac{m^2}{\mu^2}\right)
   =\frac{\beta_0\lambda(\mu^2)}{16\pi^2}\ln\left(\frac{m^2}{\Lambda^2}\right),
\notag\\
   1-\Pi(p^2=\infty,m^2)&=+\infty,
\label{eq:(B21)}
\end{align}
where we have used the dynamical scale~$\Lambda$ given by~Eq.~\eqref{eq:(B2)}
with~$d=4$.

Now, if $m^2>\Lambda^2$, Eq.~\eqref{eq:(B21)} shows that $1-\Pi(p^2,m^2)$ is
positive definite and the function~$[1-\Pi(p^2,m^2)]^{-1}$ in the integrand
of~Eq.~\eqref{eq:(B19)} does not possess any singularity; Eq.~\eqref{eq:(B19)}
is well defined. On the other hand, if the mass is small enough
as~$0<m^2<\Lambda^2$, then $1-\Pi(p^2,m^2)$ develops a simple zero in~$p^2$ and
the momentum integral~\eqref{eq:(B19)} becomes ill defined and ambiguous; this
is the renormalon precursor in the present example. Finally, in the massless
limit~$m^2\to0$, the integral~\eqref{eq:(B19)} reduces to~Eq.~\eqref{eq:(B1)}
with~Eq.~\eqref{eq:(B2)} (with~$d=4$ and rescaling the renormalization scale of
the coupling $p^2\to e^{-5/3}p^2$) and the ambiguity of the renormalon precursor
coincides with the IR renormalon ambiguity.


\begin{thebibliography}{00}

\bibitem{tHooft:1977xjm}
G.~'t Hooft,
Subnucl. Ser. \textbf{15}, 943 (1979)
doi:10.1007/978-1-4684-0991-8\_17
PRINT-77-0723 (UTRECHT).

\bibitem{Beneke:1998ui}
M.~Beneke,
Phys. Rept. \textbf{317}, 1-142 (1999)
doi:10.1016/S0370-1573(98)00130-6
[arXiv:hep-ph/9807443 [hep-ph]].

\bibitem{Argyres:2012vv}
P.~Argyres and M.~\"Unsal,
Phys. Rev. Lett. \textbf{109}, 121601 (2012)
doi:10.1103/PhysRevLett.109.121601
[arXiv:1204.1661 [hep-th]].

\bibitem{Argyres:2012ka}
P.~C.~Argyres and M.~\"Unsal,
JHEP \textbf{08}, 063 (2012)
doi:10.1007/JHEP08(2012)063
[arXiv:1206.1890 [hep-th]].

\bibitem{Dunne:2012ae}
G.~V.~Dunne and M.~\"Unsal,
JHEP \textbf{11}, 170 (2012)
doi:10.1007/JHEP11(2012)170
[arXiv:1210.2423 [hep-th]].

\bibitem{Dunne:2012zk}
G.~V.~Dunne and M.~\"Unsal,
Phys. Rev. D \textbf{87}, 025015 (2013)
doi:10.1103/PhysRevD.87.025015
[arXiv:1210.3646 [hep-th]].

\bibitem{Unsal:2007jx}
M.~\"Unsal,
Phys. Rev. D \textbf{80}, 065001 (2009)
doi:10.1103/PhysRevD.80.065001
[arXiv:0709.3269 [hep-th]].

\bibitem{Bogomolny:1980ur}
E.~Bogomolny,
Phys. Lett. B \textbf{91}, 431-435 (1980)
doi:10.1016/0370-2693(80)91014-X

\bibitem{ZinnJustin:1981dx}
J.~Zinn-Justin,
Nucl. Phys. B \textbf{192}, 125-140 (1981)
doi:10.1016/0550-3213(81)90197-8

\bibitem{Dunne:2015eaa}
G.~V.~Dunne and M.~\"Unsal,
PoS \textbf{LATTICE2015}, 010 (2016)
doi:10.22323/1.251.0010
[arXiv:1511.05977 [hep-lat]].

\bibitem{Anber:2014sda}
M.~M.~Anber and T.~Sulejmanpasic,
JHEP \textbf{01}, 139 (2015)
doi:10.1007/JHEP01(2015)139
[arXiv:1410.0121 [hep-th]].

\bibitem{Ishikawa:2019tnw}
K.~Ishikawa, O.~Morikawa, A.~Nakayama, K.~Shibata, H.~Suzuki and H.~Takaura,
PTEP \textbf{2020}, no.2, 023B10 (2020)
doi:10.1093/ptep/ptaa002
[arXiv:1908.00373 [hep-th]].

\bibitem{Ashie:2019cmy}
M.~Ashie, O.~Morikawa, H.~Suzuki, H.~Takaura and K.~Takeuchi,
PTEP \textbf{2020}, no.2, 023B01 (2020)
doi:10.1093/ptep/ptz157
[arXiv:1909.05489 [hep-th]].

\bibitem{Ishikawa:2019oga}
K.~Ishikawa, O.~Morikawa, K.~Shibata, H.~Suzuki and H.~Takaura,
PTEP \textbf{2020}, no.1, 013B01 (2020)
doi:10.1093/ptep/ptz147
[arXiv:1909.09579 [hep-th]].

\bibitem{Ishikawa:2020eht}
K.~Ishikawa, O.~Morikawa, K.~Shibata and H.~Suzuki,
arXiv:2001.07302 [hep-th].

\bibitem{Fujimori:2016ljw}
T.~Fujimori, S.~Kamata, T.~Misumi, M.~Nitta and N.~Sakai,
Phys. Rev. D \textbf{94}, no.10, 105002 (2016)
doi:10.1103/PhysRevD.94.105002
[arXiv:1607.04205 [hep-th]].

\bibitem{Fujimori:2018kqp}
T.~Fujimori, S.~Kamata, T.~Misumi, M.~Nitta and N.~Sakai,
JHEP \textbf{02}, 190 (2019)
doi:10.1007/JHEP02(2019)190
[arXiv:1810.03768 [hep-th]].

\bibitem{Kovtun:2007py}
P.~Kovtun, M.\"Unsal and L.~G.~Yaffe,
JHEP \textbf{06}, 019 (2007)
doi:10.1088/1126-6708/2007/06/019
[arXiv:hep-th/0702021 [hep-th]].

\bibitem{Unsal:2007vu}
M.~\"Unsal,
Phys. Rev. Lett. \textbf{100}, 032005 (2008)
doi:10.1103/PhysRevLett.100.032005
[arXiv:0708.1772 [hep-th]].

\bibitem{Shifman:2008ja}
M.~Shifman and M.~\"Unsal,
Phys. Rev. D \textbf{78}, 065004 (2008)
doi:10.1103/PhysRevD.78.065004
[arXiv:0802.1232 [hep-th]].

\bibitem{Unsal:2008ch}
M.~\"Unsal and L.~G.~Yaffe,
Phys. Rev. D \textbf{78}, 065035 (2008)
doi:10.1103/PhysRevD.78.065035
[arXiv:0803.0344 [hep-th]].

\bibitem{Shifman:2009tp}
M.~Shifman and M.~\"Unsal,
Phys. Lett. B \textbf{681}, 491-494 (2009)
doi:10.1016/j.physletb.2009.10.060
[arXiv:0901.3743 [hep-th]].

\bibitem{Anber:2011de}
M.~M.~Anber and E.~Poppitz,
JHEP \textbf{06}, 136 (2011)
doi:10.1007/JHEP06(2011)136
[arXiv:1105.0940 [hep-th]].

\bibitem{Unsal:2012zj}
M.~\"Unsal,
Phys. Rev. D \textbf{86}, 105012 (2012)
doi:10.1103/PhysRevD.86.105012
[arXiv:1201.6426 [hep-th]].

\bibitem{Poppitz:2012sw}
E.~Poppitz, T.~Sch\"afer and M.~\"Unsal,
JHEP \textbf{10}, 115 (2012)
doi:10.1007/JHEP10(2012)115
[arXiv:1205.0290 [hep-th]].

\bibitem{Poppitz:2012nz}
E.~Poppitz, T.~Sch\"afer and M.~\"Unsal,
JHEP \textbf{03}, 087 (2013)
doi:10.1007/JHEP03(2013)087
[arXiv:1212.1238 [hep-th]].

\bibitem{Basar:2013sza}
G.~Basar, A.~Cherman, D.~Dorigoni and M.~\"Unsal,
Phys. Rev. Lett. \textbf{111}, no.12, 121601 (2013)
doi:10.1103/PhysRevLett.111.121601
[arXiv:1306.2960 [hep-th]].

\bibitem{Poppitz:2013zqa}
E.~Poppitz and T.~Sulejmanpasic,
JHEP \textbf{09}, 128 (2013)
doi:10.1007/JHEP09(2013)128
[arXiv:1307.1317 [hep-th]].

\bibitem{Anber:2013doa}
M.~M.~Anber, S.~Collier, E.~Poppitz, S.~Strimas-Mackey and B.~Teeple,
JHEP \textbf{11}, 142 (2013)
doi:10.1007/JHEP11(2013)142
[arXiv:1310.3522 [hep-th]].

\bibitem{Cherman:2014ofa}
A.~Cherman, D.~Dorigoni and M.~\"Unsal,
JHEP \textbf{10}, 056 (2015)
doi:10.1007/JHEP10(2015)056
[arXiv:1403.1277 [hep-th]].

\bibitem{Misumi:2014raa}
T.~Misumi and T.~Kanazawa,
JHEP \textbf{06}, 181 (2014)
doi:10.1007/JHEP06(2014)181
[arXiv:1405.3113 [hep-ph]].

\bibitem{Anber:2014lba}
M.~M.~Anber, E.~Poppitz and B.~Teeple,
JHEP \textbf{09}, 040 (2014)
doi:10.1007/JHEP09(2014)040
[arXiv:1406.1199 [hep-th]].

\bibitem{Dunne:2016nmc}
G.~V.~Dunne and M.~\"Unsal,
Ann. Rev. Nucl. Part. Sci. \textbf{66}, 245-272 (2016)
doi:10.1146/annurev-nucl-102115-044755
[arXiv:1601.03414 [hep-th]].

\bibitem{Sulejmanpasic:2016llc}
T.~Sulejmanpasic,
Phys. Rev. Lett. \textbf{118}, no.1, 011601 (2017)
doi:10.1103/PhysRevLett.118.011601
[arXiv:1610.04009 [hep-th]].

\bibitem{Cherman:2016hcd}
A.~Cherman, T.~Sch\"afer and M.~\"Unsal,
Phys. Rev. Lett. \textbf{117}, no.8, 081601 (2016)
doi:10.1103/PhysRevLett.117.081601
[arXiv:1604.06108 [hep-th]].

\bibitem{Yamazaki:2017ulc}
M.~Yamazaki and K.~Yonekura,
JHEP \textbf{07}, 088 (2017)
doi:10.1007/JHEP07(2017)088
[arXiv:1704.05852 [hep-th]].

\bibitem{Aitken:2017ayq}
K.~Aitken, A.~Cherman, E.~Poppitz and L.~G.~Yaffe,
Phys. Rev. D \textbf{96}, no.9, 096022 (2017)
doi:10.1103/PhysRevD.96.096022
[arXiv:1707.08971 [hep-th]].

\bibitem{Tanizaki:2017qhf}
Y.~Tanizaki, T.~Misumi and N.~Sakai,
JHEP \textbf{12}, 056 (2017)
doi:10.1007/JHEP12(2017)056
[arXiv:1710.08923 [hep-th]].

\bibitem{Morikawa:2020agf}
O.~Morikawa and H.~Takaura,
[arXiv:2003.04759 [hep-th]].

\bibitem{Beneke:1994qe}
M.~Beneke and V.~M.~Braun,
Phys. Lett. B \textbf{348}, 513-520 (1995)
doi:10.1016/0370-2693(95)00184-M
[arXiv:hep-ph/9411229 [hep-ph]].

\bibitem{Broadhurst:1993ru}
D.~J.~Broadhurst and A.~Kataev,
Phys. Lett. B \textbf{315}, 179-187 (1993)
doi:10.1016/0370-2693(93)90177-J
[arXiv:hep-ph/9308274 [hep-ph]].

\bibitem{Ball:1995ni}
P.~Ball, M.~Beneke and V.~M.~Braun,
Nucl. Phys. B \textbf{452}, 563-625 (1995)
doi:10.1016/0550-3213(95)00392-6
[arXiv:hep-ph/9502300 [hep-ph]].

\bibitem{Gross:1982at}
D.~J.~Gross and Y.~Kitazawa,
Nucl. Phys. B \textbf{206}, 440-472 (1982)
doi:10.1016/0550-3213(82)90278-4

\bibitem{Gross:1980br}
D.~J.~Gross, R.~D.~Pisarski and L.~G.~Yaffe,
Rev. Mod. Phys. \textbf{53}, 43 (1981)
doi:10.1103/RevModPhys.53.43

\bibitem{Kapusta:2006pm}
J.~Kapusta and C.~Gale,
``Finite-temperature field theory: Principles and applications,''
doi:10.1017/CBO9780511535130

\bibitem{Novikov:1984rf}
V.~Novikov, M.~A.~Shifman, A.~Vainshtein and V.~I.~Zakharov,
Yad. Fiz. \textbf{41}, 1063-1079 (1985)
doi:10.1016/0550-3213(85)90087-2

\end{thebibliography}
\end{document}